\documentclass[aps,prl,twocolumn,showpacs,superscriptaddress,groupedaddress]{revtex4-1}  
        \clearpage
\usepackage{graphicx}  
\usepackage{dcolumn}   
\usepackage{bm}        
\usepackage{amssymb}   
\usepackage{amsmath}
\usepackage{xspace}
\usepackage[caption = false]{subfig}
\usepackage{ulem} 

\usepackage{times}
\usepackage{color}
\usepackage{ulem}

\newcommand{\be}{\begin{equation}}
\newcommand{\ee}{\end{equation}}

\begin{document} 
\title{On the maximal efficiency of the collisional Penrose process }


\author
{Elly Leiderschneider and Tsvi Piran }
\affiliation{ Racah Institute of Physics, The Hebrew University of Jerusalem, Jerusalem 91904, Israel }
\date{\today}

\begin{abstract}
{\bf Abstract:}  

The center of mass (CM) energy  in a collisional Penrose process -  a collision taking place within the ergosphere of a Kerr black hole - can diverge under suitable extreme conditions (maximal Kerr, near horizon collision and suitable impact parameters). We present an analytic expression for the CM energy, refining expressions given in the literature. Even though the CM energy diverges, we show that the maximal energy attained by a particle that escapes the black hole's gravitational pull and reaches infinity is modest.
We obtain an analytic expression for the energy of an escaping particle resulting from a collisional Penrose process, and apply it  to derive the maximal energy and the maximal efficiency for several physical scenarios:  pair annihilation, Compton scattering, and the elastic scattering of two massive particles. In all physically reasonable cases (in which the incident particles initially fall from infinity towards the black hole) the maximal energy  (and the corresponding efficiency) are  only one order of magnitude larger than the rest mass energy of the incident particles. The maximal efficiency found is $\approx 13.92$ and it is obtained for the scattering of an outgoing massless particle by a massive particle. 

\end{abstract}
\maketitle

\section{1. Introduction}
\label{sec:intro}

\begin{figure*}[]
\includegraphics[width=43mm]{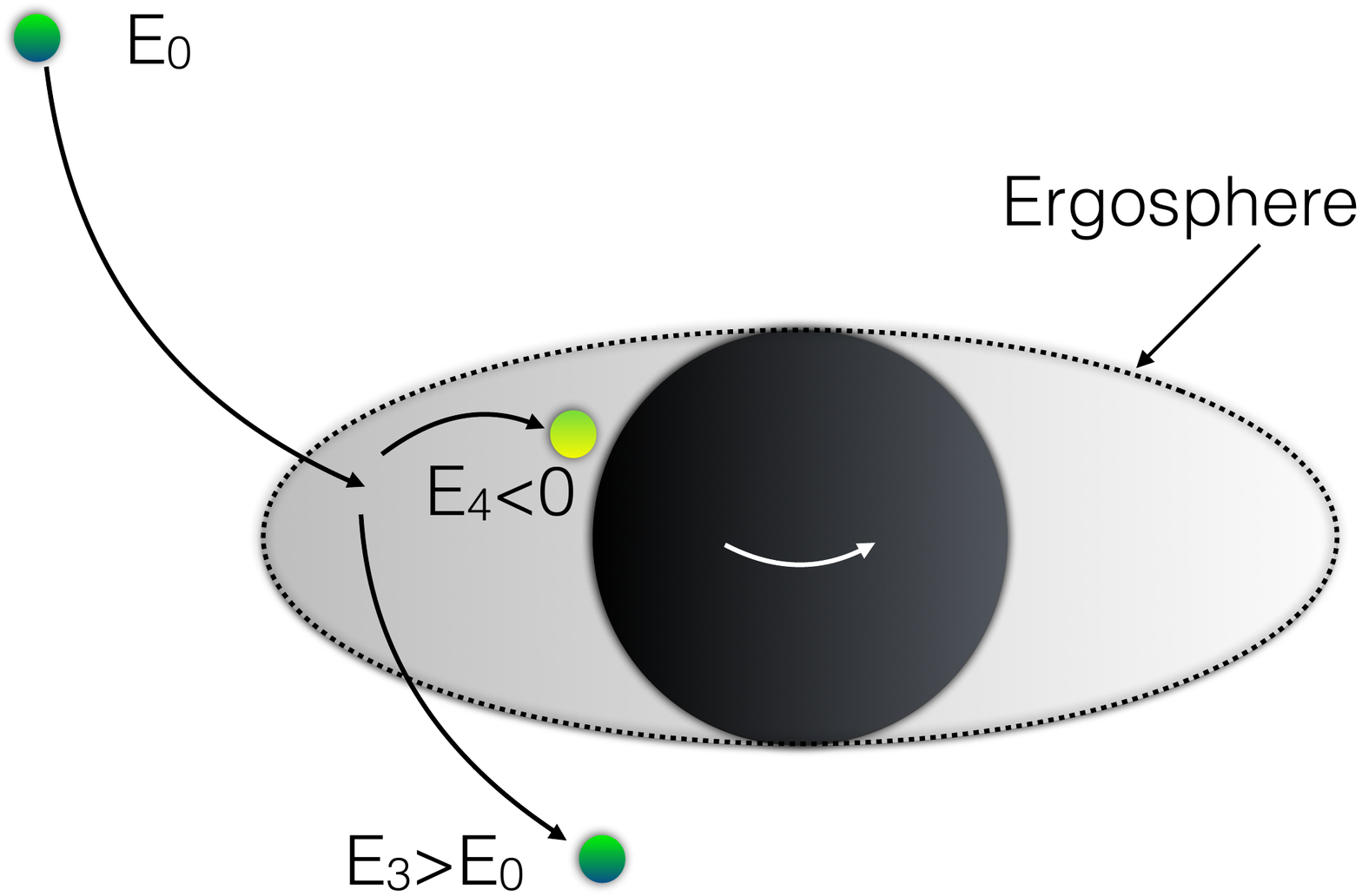}
\includegraphics[width=43mm]{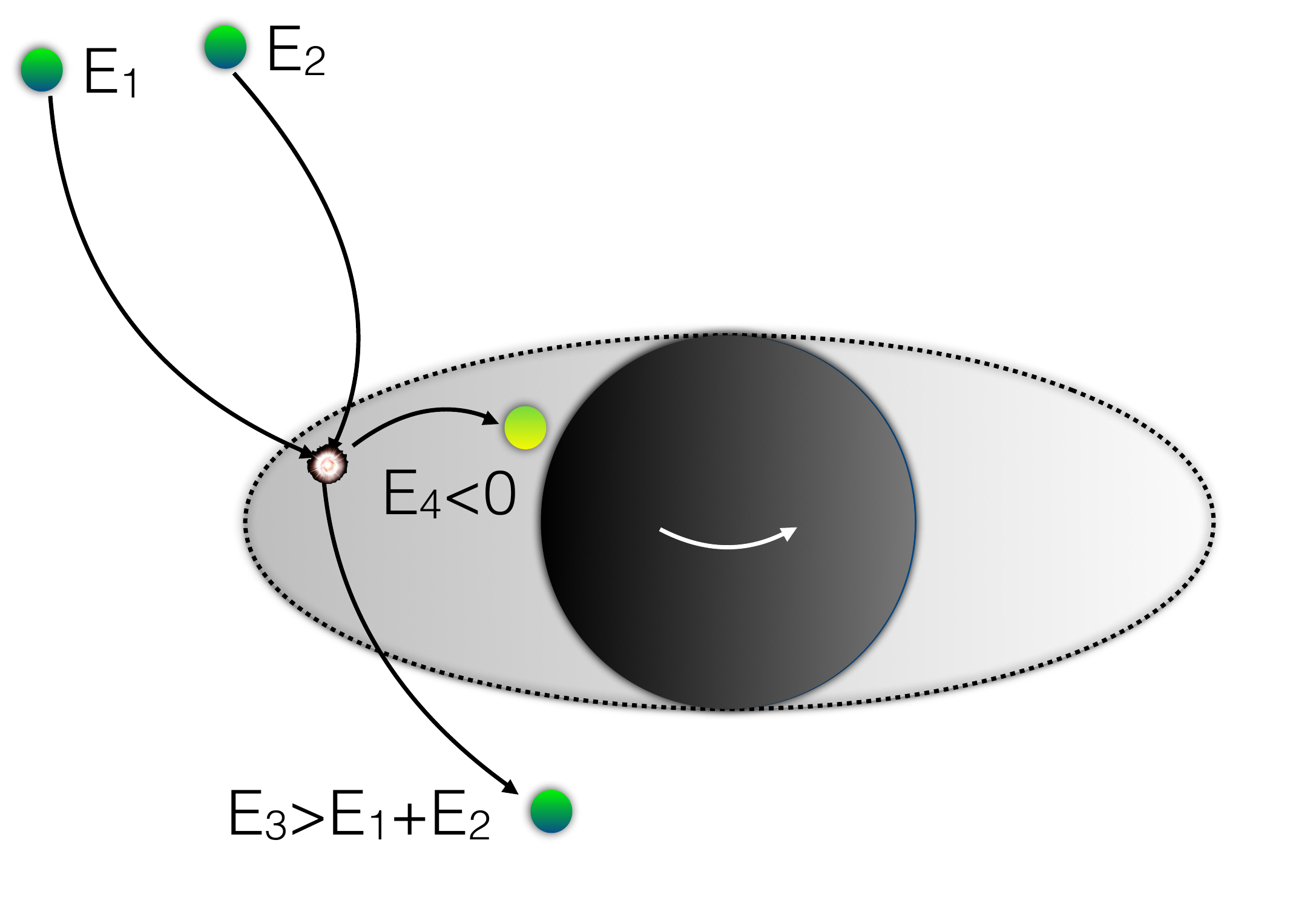}
\includegraphics[width=43mm]{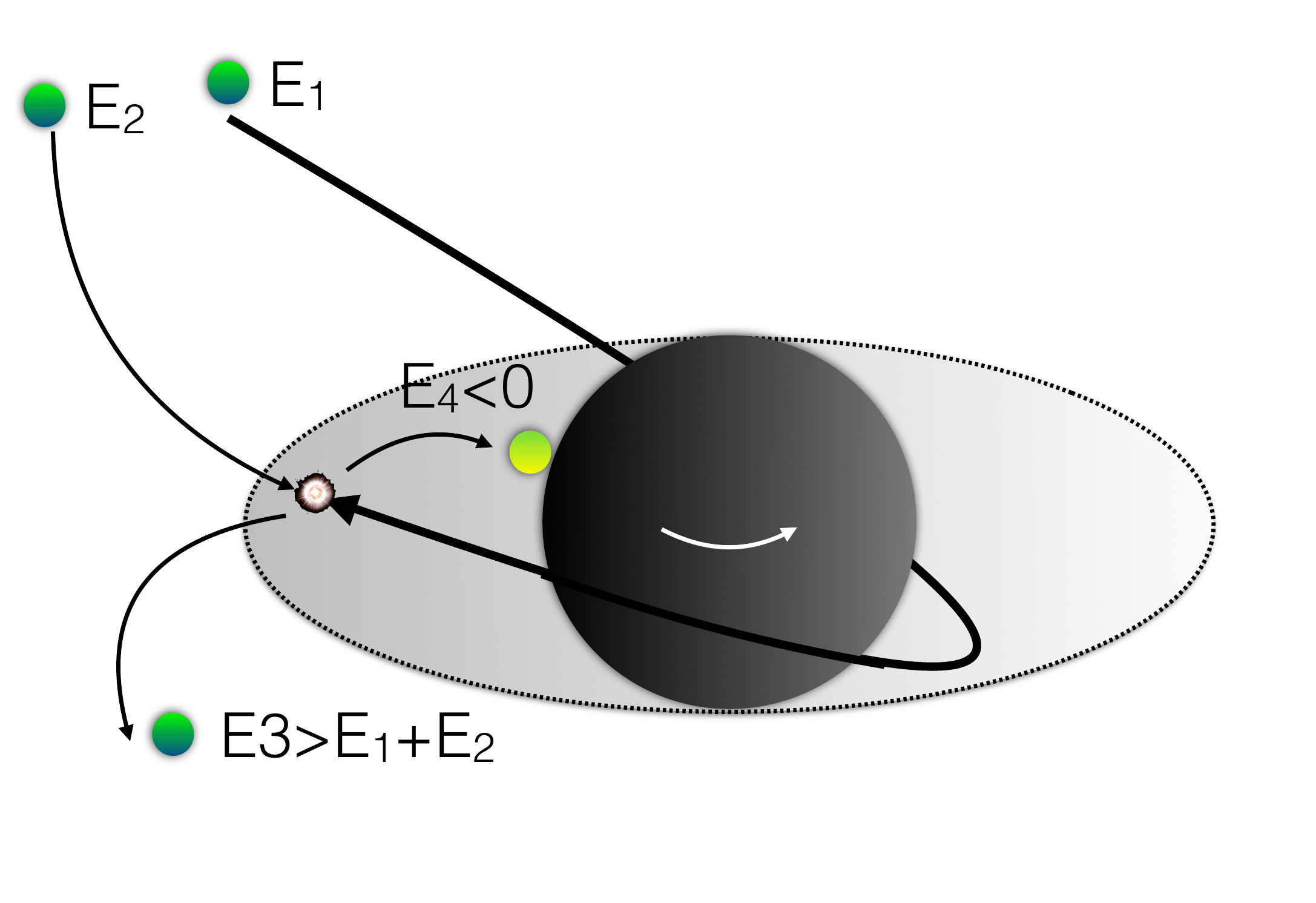}
\includegraphics[width=43mm]{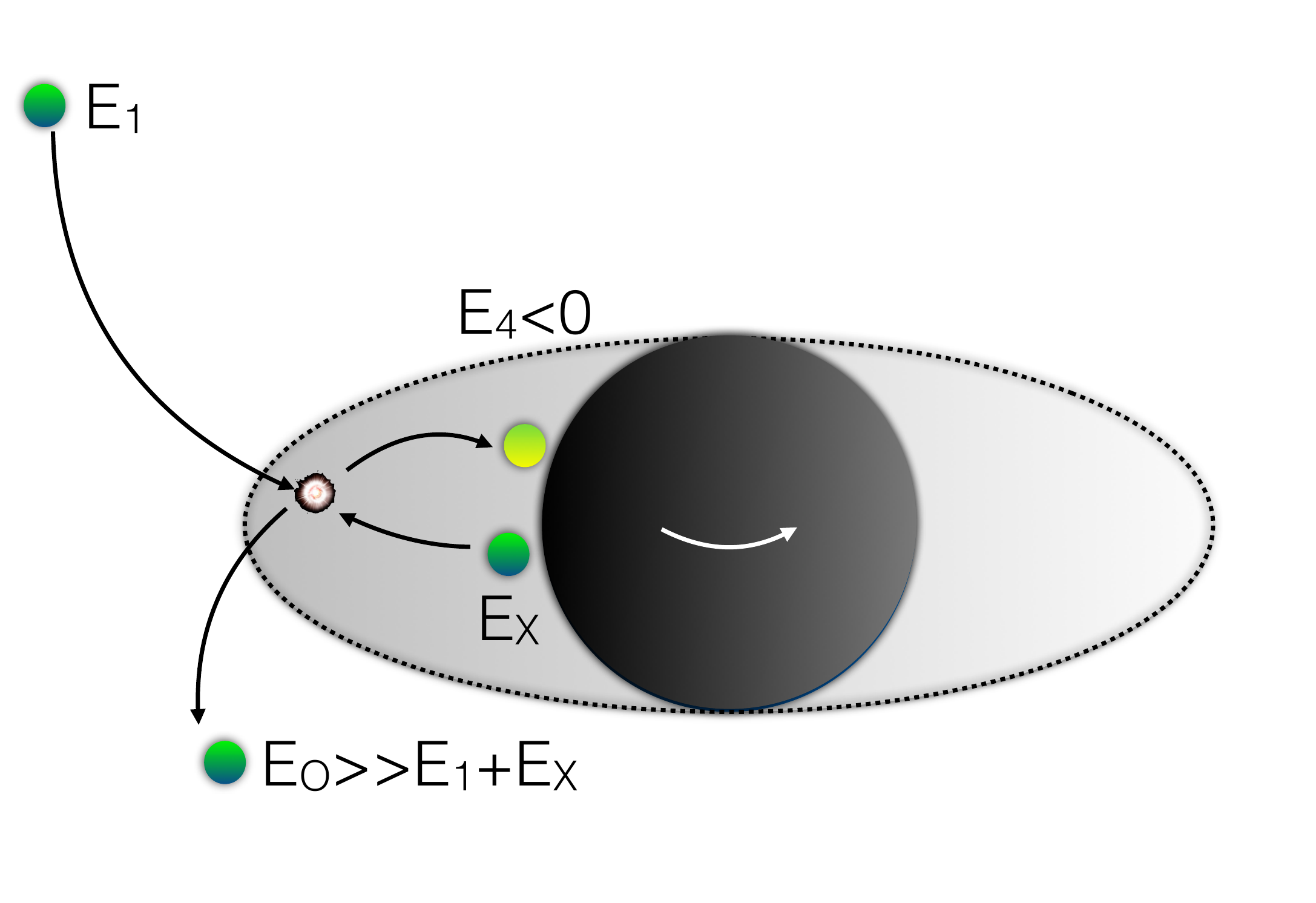}
\caption{ Left: The original Penrose process  -  a particle disintegrates inside the ergosphere. One of the resulting particles possesses negative energy, and falls into the black hole. The other one escapes to infinity with more energy than that of the initial particle   \cite{Penrose69}. Center left: A collisional Penrose process. Two infalling particles collide in the ergosphere   \cite{Piran+75}.
Center right: One of the initially infalling particles turns around the black hole and collides with a second infalling particle while it is outgoing   \cite{Schnittman14}. Right: An ingoing particle (1) collides with an outgoing particle $_X$ emerging from the black hole   \cite{Berti+14}. }
\label{fig:penrose}
\end{figure*}

The collisional Penrose process was suggested 
by  Piran, Shaham and Katz \cite{Piran+75} once it was realized that the original Penrose process \cite{Penrose69} is inefficient \cite{Bardeen+72,Wald74, KovetzPiran75}.  In the original  process \cite{Penrose69} a particle disintegrates in the ergosphere of a Kerr black hole into two: a negative energy particle that plunges into the black hole, and a second particle that escapes to infinity with energy larger than that of the original (see Fig.  \ref{fig:penrose}). The energy gain arises, of course, from the rotational energy of the Kerr black hole that absorbs the negative energy particle.
However,    in order that the positive energy particle escapes to infinity, a significant fraction of the original infalling particle's rest mass must be converted to energy in its rest frame \cite{Bardeen+72,Wald74, KovetzPiran75}. The energy gain from the black hole is  insignificant compared with the energy conversion in the particle's rest frame, making this process somewhat "uninteresting" from an astrophysical or technological \footnote{It was speculated that an advanced civilization could power itself by dumping its garbage on a Kerr black hole and extracting, via the Penrose process, the black hole's rotational energy}  point of view.

In the collisional Penrose process \cite{Piran+75}, two  particles  collide in the ergosphere (see Fig.  \ref{fig:penrose}), producing a negative energy particle, as well as a positive energy one whose energy is larger that the sum of the  initial energies of the colliding particles. 
Recent interest in this process arose when Banados et al.   \cite{Banados+09} noticed that the energy in the CM frame diverges in some collisions that take place  near the horizon of  an extreme  Kerr  black hole. It was suggested that this infinite energy could be utilized to accelerate particles to extremely large energies or produce exotic massive particles that cannot be produced otherwise. However, while the energy in the CM is extremely large, this study   \cite{Banados+09} and many subsequent ones ignored the requirement that the resulting energetic particle escapes to infinity.   Piran and Shaham  \cite{PiranShaham77} have shown that this requirement  imposes stringent conditions on the dynamics, and is critical when estimating the maximal possible energy gain in this process. 
They provided  implicit estimates of the energies of the escaping particles under various conditions.


Detailed numerical studies   \cite{Bejger+12} have refined some of the earlier analysis of Penrose collisions. These studies have shown that the maximal energy gain is, at most, a few times the energy of the infalling particles and the efficiency is modest.  When two infalling particles collide near the horizon, their CM has an enormous negative radial momentum. Hence, most of the particles produced in such collisions will also move inwards and fall into the black hole. Only a small fraction will escape, and those have to climb out of the black hole's potential well and reach infinity only with modest energies. 

While both particles are initially infalling from infinity, it is possible that one of those has turned around the black hole and is on an outgoing trajectory   \cite{Schnittman14}. This reduces somewhat the magnitude of the negative radial momentum  (but it is insufficient to change its sign), and as such the maximal possible energy in such a collision is larger than that of the case where both particles are infalling. Both \cite{Bejger+12} and  \cite{Schnittman14} used numerical methods to explore fully-equatorial collisions, in which both incident and ejected particles are constrained to move within the equatorial plane. As we show later, the numerical limit obtained by \cite{Bejger+12} can be surpassed by a non-equatorial collision, while that obtained by \cite{Schnittman14} is valid also for the general case. 

Here we consider general collisions of particles infalling from infinity that result in an escaping particle.  We 
obtain a general analytic formula for the maximal energy of the escaping particle. The formula is also valid outside of the equatorial plane. It can also be easily generalized to the case of a collision involving an arbitrary number of incident or ejected particles. 
We explore the implications of this formula for several possible scenarios: pair annihilation, the elastic scattering of two massive particles, and the scattering of a photon by a massive particle  (e.g. Compton  scattering). We are interested in the maximal energy of the escaping particles and in the efficiency of the process, defined as the ratio of the energy of the escaping particle to the sum of the energies of the incident particles, and we focus on the maximal efficiency or, correspondingly, on the maximal energy gain. 

As  orbits in the Kerr metric are essential for our study, we begin in \S 2 \ref{sec:orbits} with a quick recapitulation of the relevant properties of these orbits. We continue in  \S 3 \ref{sec:general}
with an examination of collisions, and derive the formula describing the energy of a particle produced by a general collision. We first discuss the general case, then provide simplified results for collisions wherein all particles are constrained to the equatorial plane.  We examine in \S 4  \ref{sec:scenarios} the implications for annihilation, elastic scattering and Compton - like scattering for both equatorial and non-equatorial collisions.  In all cases we discuss both the case in which the two incident particles are ingoing, and the case in which one of them has turned around the black hole and is outgoing.                                                                                                                                                                                      
We summarize our results in 
\S  5 \ref{sec:conclusions}, and discuss some implications to astrophysical or ``technological" applications of this mechanism.

\section{2. Orbits in Kerr}
\label{sec:orbits} 

We describe the Kerr metric using the Boyer - Lindquist coordinates. The Kerr solution is characterized by two parameters: the mass, $M$, and the specific angular momentum,  $a$. Without loss of generality we set $M=1$. 

A geodesic is described by four parameters: $E$, $L$ $Q$ and  $m$, the particle's energy, axial angular momentum, Carter constant and mass. The latter is taken to be 1 for massive particles and 0 for photons.  Note that as nowhere we compare the particles' masses to the black hole's mass, it is possible to set both to unity.  This sets $M$ as a unit of length, and $m$ as a unit of energy. We define the particle's impact parameter as $b \equiv L/E$,  and will use both $b$ and $L$ in the following.   

The radial  momentum is given in terms of these quantities as
\begin{equation}
\label{Eq:pr}
p^{r}=\epsilon\frac{\sqrt{V (r)}}{\Sigma}\ , 
\end{equation}
where
\begin{equation}
\label{Eq:vr}
V (r)=E^{2}r^{4} - 
 (L^2 - a^{2}E^{2} )r^{2} + 2 (L - aE)^{2}r -   (m^2 r^ 2 + Q) \Delta  \  ,
\end{equation}
and we used the common definitions: 
\begin{equation} 
\Delta \equiv  (r^2  -  2 M r + a^2)  \ \ ;  \ \ \Sigma = r^2+a^2 cos^2\theta\ . 
\end{equation}
The factor $\epsilon=\pm1 $ reflects the sign of the radial momentum  ( -  for ingoing, + for outgoing).
The polar  ($\theta$) momentum is given by
\begin{equation}
p^{\theta}=\frac{\epsilon^{\theta}}{\Sigma}\sqrt{Q - \cos^{2}\theta \left[\frac{L^{2}}{\sin^{2}\theta} - a^{2} (E^{2} - m^{2}) \right]} \ ,
\label{Eq:ptheta}
\end{equation}
where $\epsilon^{\theta} = \pm 1$ depending on the direction of the polar motion of the particle. Eq.  \ref{Eq:ptheta} implies a minimal value for $Q$,
\begin{equation}
Q\geq \cos^2\theta[\frac{L^2}{\sin^2\theta} - a^2 (E^2 - m^2)] .
\label{Eq:Qmin}
\end{equation}

\begin{figure*}
\includegraphics[width=85mm]{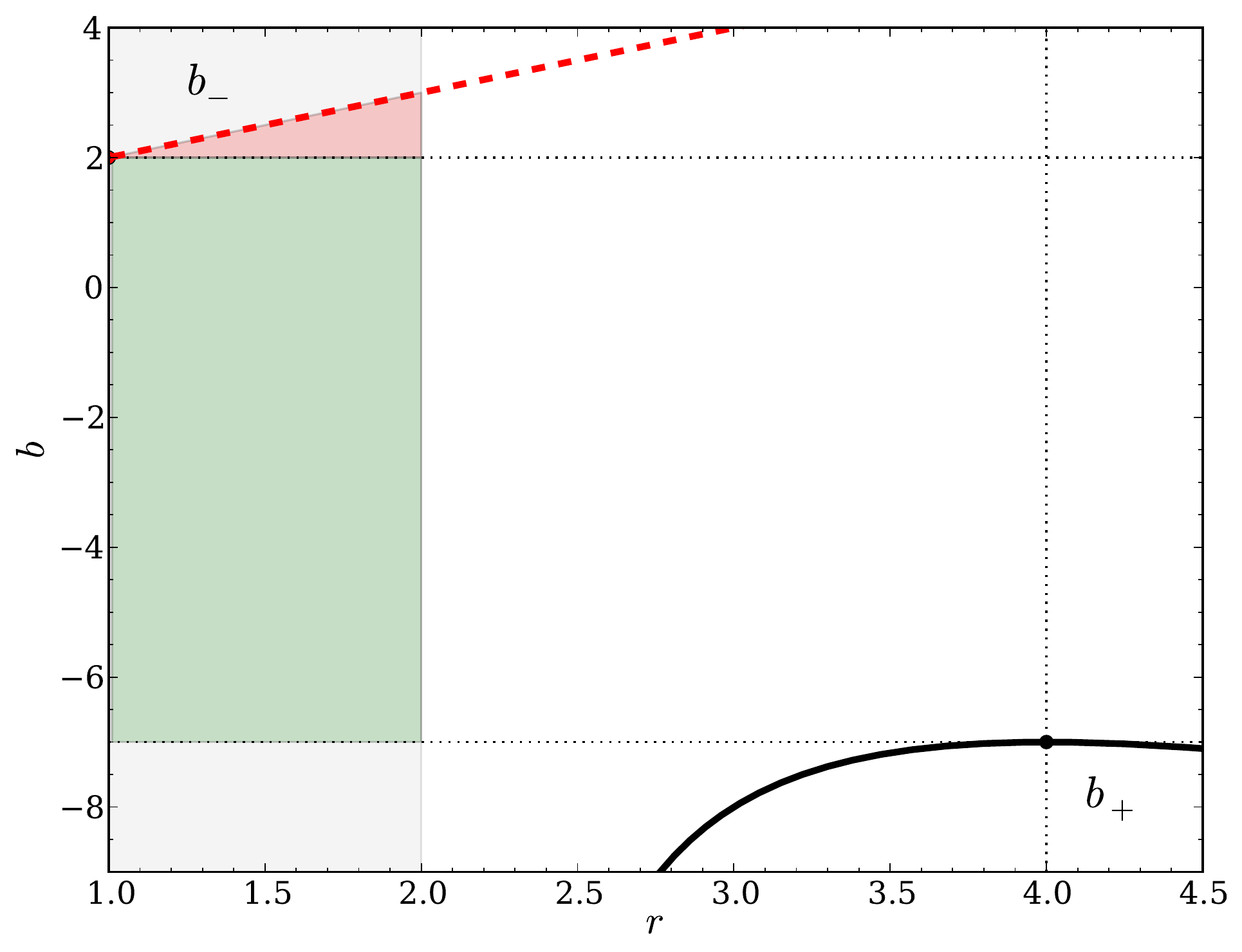}
\caption{From \cite{Bejger+12}: The photon turning point diagram for $a=1$. The red region denotes the initially ingoing photons that can escape to infinity.}
\label{fig:b_a1}
\end{figure*}

A particle has a radial turning point  where $V (r)=0$.  Particularly important are the photon's equatorial turning points, which take place at
\begin{equation} 
b_{\pm} (r)=\frac{2a\pm\sqrt{r^{4} - 2r^{3}+a^{2}r^{2}}}{2 - r}.
\label{Eq:bturn}
\end{equation} 
A turning point diagram in the $ (r,b)$ plane is a union of two
graphs (See Fig. \ref{fig:b_a1}). The upper branch has a minimum $b_{u}$ at
$r=r_{u}$, and the lower one has a  maximum of $b_{l}$
at $r=r_{l}$. Only photons with impact parameters $b_{l}<b<b_{u}$ can freely pass between the horizon and infinity.
For a general $a$, these are found
to be  (see   \cite{Chandrasekhar})
\begin{equation} 
 b_{u}=y_{u} - a, \quad  r_{u}=3 (1 - 2a/y_{u}) ,
\end{equation} 
\label{Eq:bmax}
where $y_{u}= - 6\cos[ ({2\pi+\cos^{ - 1}a})/{3}]$,
and
\begin{equation} 
 \quad b_{l}=y_{l}+a, \quad r_{l}=3 (1+2a/y_{l})
\label{Eq:bmin}
\end{equation}
where $ y_{l}=6\cos[{ (\cos^{ - 1}a)}/{3}]$.
For $a=1$, which we use in the following, $b_{u}=2$, $r_{u}=1$
and $b_{l}= - 7$, $r_l=4$ .

We will be interested in photons produced in a near-horizon collision that escape to infinity. An initially outgoing photon must have $b>b_{l}$ in order to escape to infinity, while an initially
ingoing one must have $b\geq b_{u}$ and be produced at $r\geq r_{u}$  (since
it needs to bounce back off the potential). 

The innermost turning point is located at $r=r_{u}$, which is always larger than the horizon , $r_H\equiv 1+\sqrt{1 - a^2}$.  As $a \rightarrow 1$ both  $r_{u}$ and $r_H$ approach 1,
with  $r_{u} > r_H$. In the limit $r_{u}=r_H=1$,  but  the last inequality still holds and the
two radii don't coincide. This is due to a coordinate singularity at $r=1$ for $a=1$  (see 
  \cite{Bardeen+72} for details.).  An initially ingoing particle can therefore approach closest to the horizon with $a=1$.  If the turning point is 
 further from the horizon, the maximum efficiency of the process is lower. Hence, while the problem is soluble with a general $a$,  for the sake of clarity we simplify the analysis and 
set $a=1$ in the rest of the text.

For $a=1$ , the upper branch of the turning point diagram reduces to $b_{-}= r+1$.
An initially infalling equatorial photon produced at $r$  can therefore escape to infinity if its impact parameter 
$b$ satisfies(see the red region in Fig. \ref{fig:b_a1})
\begin{equation}
2  \leq b \leq r+1 \ .
\label{Eq:escape}
\end{equation}
The first inequality arises because  the photon has to bounce back off its turning point. The second arises so that the particle is in the allowed region where $V (r)\geq0$.

To approach nearest to the horizon before being deflected,
a massive particle or a photon must have  $b=2$. We denote the value $b=2$ as the critical impact parameter. A critical particle will have $b=2$ and a subcritical one will have $b<2$. 

So far we discussed equatorial turning points. When moving out of the equatorial plane, a non - vanishing Carter constant, $Q$, that determines the polar motion, can also influence 
 the particle's turning point diagram in the radial direction. 
 
Even for $Q\ne 0$  the minimal turning point, $r_{u}$, remains 1  (and correspondingly $b_{u}=2$)
 for $Q\leq 2 E^2$ (which sets a lower bound, $\sin\theta\geq\sqrt{2/3}$) for massive particles, or for $Q\leq 3 E^2$ (which sets a lower bound, $\sin\theta\geq\sqrt{3} - 1$) for photons.  At $\sin\theta < \sqrt{2/3}$ for massive particles, or $\sin\theta < \sqrt{3} - 1$ for  photons, $b_{u}$ decreases and $r_{u}$ increases.

\section{3. The collisional Penrose process}
\label{sec:general}

We consider a collision between two incident particles  (1,2) that results in two ejected particles  (3,4). We begin by finding a general expression for the  energy of the escaping particle  produced in such a collision.
As we show later, this expression holds for an arbitrary number of incident or ejected particles.
We denote by (3) the particle that escapes to infinity.  If there is energy gain, $E_3 > E_1 + E_2$, then $E_4 < 0 $, and particle (4)  (which could effectively be a composite particle) must plunge into the black hole.

Energy and momentum conservations imply:
\begin{equation} 
E_{tot}\equiv E_{1}+E_{2}= E_{3} +E_{4} \ ,
\label{Eq:E}\end{equation} 
\begin{equation} 
L_{tot}\equiv b_1 E_{1}+ b_2 E_{2}= b_3 E_{3} +b_4 E_{4}  \ , 
\label{Eq:L}\end{equation} 
\begin{equation}
p_{tot}^{r} \equiv  \epsilon_1 p^r_1+\epsilon_2 p^r_2 =\epsilon_3 p^r_3+\epsilon_4 p^r_4 \ ,
\label{Eq:prC}
\end{equation}
 and 
\begin{equation}
p_{tot}^{\theta}\equiv \epsilon^{\theta}_1  p^\theta_1+ \epsilon^{\theta} _2 p^\theta_2 = \epsilon^{\theta} _3 p^\theta_3+ \epsilon^{\theta} _4 p^\theta_4 \ .
\label{Eq:pthetaC}
\end{equation}
\begin{widetext}

\subsection{The CM energy} 

The CM system has an effective mass: 
\begin{equation}
\label{Eq:MCM}
M^2_{CM}= \frac{E_{tot}^2\big [r^4 + \left (a^2 - b_{tot}^2\right) r^2 +\Delta  \cos ^2\theta \left (b_{tot}^2 /\sin^{2}\theta - a^2\right)+2 r  (b_{tot} - a)^2 \big]  - \Sigma ^2 \big [ (p_{tot}^r)^2+\Delta 
    (p_{tot}^\theta)^2\big]}{\Delta  \Sigma } \ .
\end{equation}
\end{widetext}
We determine the CM energy of a collision taking place very close to the black hole at $r=1+ \tilde \epsilon$, for a collision  
between two particles falling from rest at  infinity($E_1=E_2=m\equiv 1$) towards an extreme black hole ($a=1$).   We consider first the cases when one of the particles (2) has $b_2 <2$, and is falling towards the black hole, while the other particle (1) has $b_1 \geq 2$, and can be either ingoing or outgoing (if it has passed its turning point). 

An infalling  particle  with $E=m$, $Q$ and  $b=2+\chi$ will have a turning point at $r=1+ \chi/(2-\sqrt{2+\tilde Q})$, where $\tilde Q \equiv Q/m^2$. Thus $ \tilde \epsilon >  \chi/(2-\sqrt{2+\tilde Q})$, i.e., $\tilde \epsilon$ provides an upper bound on $\chi$. 
Using this fact and expressing $\chi$ in terms of this upper limit,  we find the maximal energy as a function of the collision point. This maximal energy diverges as
$\tilde \epsilon \rightarrow 0$  (i.e. when the collision point approaches the horizon). 
We set $\chi=\delta \tilde \epsilon$, for $\delta\leq 2-\sqrt{2+\tilde Q}$, and upon taking the limit $\tilde \epsilon \rightarrow 0$ we find the leading term
\begin{equation}
M_{CM} ^2= \frac{2m^2 (2-b_2)(2-\delta+\epsilon_1 \sqrt{2-\tilde Q_1-4\delta+\delta^2})}{\tilde \epsilon [1 + \cos^2(\theta)]}+O(1)
\label{Eq:Mcmlimit}
\end{equation}

\begin{figure*}
\includegraphics[width=85mm]{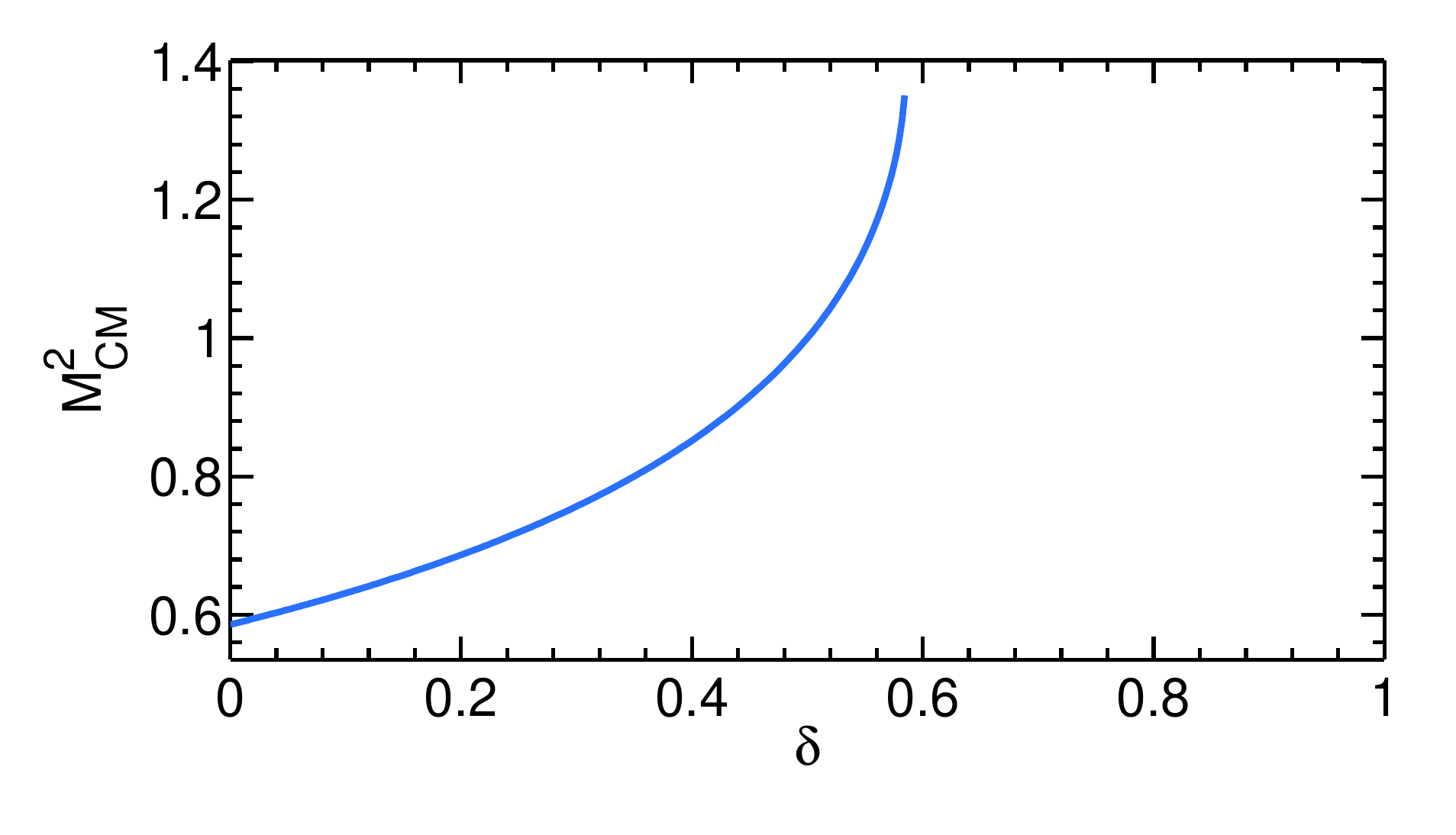}
\caption{ The prefactor of the $1/\epsilon$ term of the CM energy in Eq. \ref{Eq:Mcmlimit} for $Q_1=0 $, on the equatorial plane and for an ingoing particle (1), as a function of $\delta$. }
\label{fig:MCM}
\end{figure*}

Choosing the $\delta$ that maximizes the expression for $\epsilon_1=-1$ we obtain for an ingoing particle (1) 
\begin{equation} 
M^2_{CM} =   \frac{2 m^2(2-b_2) \sqrt{2 + \tilde Q_1}}
{\tilde \epsilon [1 + \cos^2(\theta)]}  +O(1) \  .
\label{M2max1a}
\end{equation}
If particle (1) is outgoing, the energy is : 
\begin{equation} 
M^2_{CM} =  \frac{2 m^2 (2-b_2)(2+ \sqrt{2 - \tilde Q_1})}
{\tilde \epsilon [1 + \cos^2(\theta)]}  +O(1) \ ,
\label{M2max2a}
\end{equation}
In both cases (of ingoing or outgoing particle (1)) the maximal CM mass is attained when the collision takes place in the equatorial plane. 
However, if particle (1) is ingoing, the maximal CM mass  is attained for $b_1=2$, $\tilde Q_1 \ne 0$ (This means that, while the collision is on the equatorial plane, the particle is not moving within the equatorial plane but just crosses it) and the maximum is attained for $\tilde Q_1=2$.
For an outgoing particle (1)  the maximum is attained for $\tilde Q=0$. 

For motion within the equatorial plane and $\tilde Q=0$, the expression, Eq. \ref{M2max2a}, for an outgoing particle (1) agrees  with \cite{Schnittman14} (see Figure 4 of \cite{Schnittman14}). For an ingoing particle (1)  Eq. \ref{Eq:Mcmlimit}  
agrees with the results of \cite{Schnittman14} when substituting $\delta=0$ (i.e. when assuming that the maximal energy is always at $b_1=2$). However the maximum he obtains is smaller than the maximum we find in Eq. \ref{M2max1a} (see Fig \ref{fig:MCM}). These expressions  (and in particular Eq. \ref{M2max1a}) are inconsistent with Eq. 15 of  \cite{Banados+09},  who neglected some terms when taking the limit $r\rightarrow 1$.

If both $b_1<2$ and $b_2<2$, both particles are ingoing. Now for a collision taking place at $r=1+\tilde \epsilon$, we find, in the limit $\tilde \epsilon\rightarrow 0$:
\begin{eqnarray} 
M^2_{CM} &=&  \frac{m^2}{1 + \cos^2(\theta)}   \bigg(  \frac{(2-b_2) {(2 + \tilde Q_1)}} 
{(2-b_1)  }  +  \nonumber  \\
&& \frac{(2-b_1) (2 + \tilde Q_2)}
{(2-b_2)  } \bigg) + O(1)  \ .
\label{M2max3}
\end{eqnarray}
This expression diverges if either $b_1$ or $b_2 \rightarrow 2$ and for $\tilde Q=0$ it resembles Eq. 15 of \cite{Banados+09}.
The absolute maximum is attained  on the equatorial plane but with $\tilde Q=2$.


\subsection{The energy of an outgoing particle}

A collision that involves two outgoing particles will have $8$ degrees of freedom. These can be given by the particles' constants: $\{E_{3,4} , m_{3,4} , L_{3,4} , Q_{3,4}\}$. The conservation of the 4 - momentum will eliminate 4 of these, leaving us with 4 independent variables. By choosing them correctly, we can use the conservation of radial momentum  (Eq.  \ref{Eq:prC}) to express $E_3$  as a function of those variables. 

This approach should be contrasted with earlier approaches   \cite{PiranShaham77,Bejger+12,Schnittman14},
in which the four degrees of freedom were chosen as the two angles of scattering in the CM frame and the masses $m_{3,4}$, also in the CM frame. This previous approach is more suitable for numerical simulations in which one 
knows the angular dependence of the cross section of a given process, and one explores the whole distribution of resulting energies\cite{Schnittman15}.

For the escaping particle (3), we define
\begin{equation} 
Q_3 \equiv \lambda_3^{2} E_3^{2}\geq \lambda_{3,0} ^2 E_3 ^2
\label{Eq:Qph}\end{equation} 
where 
\begin{equation}
\lambda^{2} (\theta) \equiv \cos^{2}\theta[{b^{2}}/ (\sin^{2}\theta) - 1+\alpha^2]+ (\tilde p^\theta) ^2 \ ,
\label{Eq:MuDef}\end{equation}
\begin{equation}
\lambda_0 ^{2} (\theta) \equiv \cos^{2}\theta[{b^{2}}/ (\sin^{2}\theta) - 1+\alpha^2] \ ,
\label{Eq:Mu0Def}\end{equation}
\begin{equation}
\tilde p^\theta=\frac{\Sigma p^\theta}{E} \ , 
\end{equation}
and
\begin{equation}
\alpha=\frac{m}{E} \ . 
\end{equation}

By defining  $L=b E$, $m\equiv \alpha E$ and $Q \equiv \lambda^2 E^2$, the radial momentum of a given particle (Eq.  \ref{Eq:pr}) is made manifestly proportional to its energy:
\begin{equation} 
 p^{r}_{3}
=\epsilon_3 \nu_3 E_3\ ,
\label{Eq:pr3q}\end{equation} 
where we have defined the quantity
\begin{equation} 
\nu^2 \equiv\frac{ {r^4 -  ({b^{2} - 1){r^{2}}+2 (b - 1)^{2}}{r} -  ( {\lambda^{2} } + 
\alpha^2 r^2) }\Delta}{\Sigma^2} \ . 
\label{Eq:pr3q}\end{equation}

The Carter constant is not conserved in the collision, but
by using the equation of polar momentum conservation, $Q_{4}$
can  be expressed as
\begin{equation} 
Q_{4}= (\Sigma p_{tot}^{\theta} -\tilde p_3 ^\theta E_3)^{2}+\cos^{2}\theta[\frac{(L_{tot}-b_3 E_3)^{2}}{\sin^{2}\theta} -  ((E_{tot}-E_3)^{2} - m_{4}^{2})] \ .
\label{Eq:Q4}\end{equation} 

We also define 
the effective Carter constant
\begin{equation}
Q_{N} \equiv  (\Sigma p_{tot}^{\theta})^2+\cos^2\theta (\frac{L_{tot}^{2}}{\sin^2\theta} - E_{tot}^{2}+m_4 ^2) \ .
\end{equation}

To summarize, we have chosen $\{b_3  , \tilde p_3 ^\theta ,\alpha_3 , m_4\}$ as our independent variables. $E_4$, $L_4$ and $Q_4$ are eliminated using  the corresponding conservation laws -    Eq.  \ref{Eq:E},  \ref{Eq:L} and  \ref{Eq:pthetaC}, respectively.
The radial momentum equation(Eq.  \ref{Eq:prC}) then becomes a quadratic equation in $E_3$, and can therefore be solved exactly. In fact, it turns out to be a \textit{linear} equation in the case where particle (3) is massless  ($m_3=\alpha_3 = 0$). 
Solving Eq.  \ref{Eq:prC} for $E_3$, we obtain for a massless particle (3):
\begin{widetext}
\begin{eqnarray}
E_{3}&& (E_{tot},L_{tot} ,p_{tot} ^r, p_{tot} ^\theta,b_3,\tilde p^\theta _3, m_4 ,r,\theta)
\nonumber \\
 &&= \frac{E_{tot} \Sigma \Delta(M_{CM}^2-m_4 ^2)}{ {2 ({E_{tot}^{2}} r^4 -  ({b_3 L_{tot}E_{tot} - E_{tot}^2}){r^2}+2 ({ (b_3 E_{tot} - E_{tot}) (L_{tot} - E_{tot})}){r} - {Q_{D}}\Delta -  } \Sigma p_{tot}^{r}\cdot \Sigma \epsilon_3 \nu_3 E_{tot})  }
\nonumber \\ && = \frac{ E_{tot}
[ {{E_{tot}^{2}r^4} -  ({L_{tot}^{2} - E_{tot}^{2}}){r^{2}}+
2 { (L_{tot} - E_{tot})^{2}}{r}} -  (m_4 ^2 r^2+ Q_N)\Delta  -   (\Sigma p_{tot}^r)^2 ] 
}{  {2 ({E_{tot}^{2}} r^4 -  ({b_3 L_{tot}E_{tot} - E_{tot}^2}){r^2}+2 ({ (b_3 E_{tot} - E_{tot}) (L_{tot} - E_{tot})}){r} - {Q_{D}}\Delta -  } \Sigma p_{tot}^{r}\cdot \Sigma \epsilon_3 \nu_3 E_{tot})  }  \ , 
\label{Eq:E3}
\end{eqnarray} 
\end{widetext}
where we have defined the auxiliary quantity 
$Q_D$:
\begin{equation} 
Q_D\equiv {\tilde p}_3 ^\theta E_{tot} p_{tot} ^\theta+ \cos^2\theta (\frac{b_3 E_{tot}L_{tot}}{\sin^2\theta} - E_{tot}^2) \ . 
\end{equation}
{The numerator of Eq. \ref{Eq:E3},  $\Sigma \Delta(M_{CM}^2-m_4 ^2)$, is the first indication that the energy of the particle as measured at infinity remains finite, despite the fact that  $M_{CM}$  may diverge on the horizon.}

Eq.  \ref{Eq:E3}  depends only on the \textit{total} properties of the incident particles (total energy, total angular momentum, total polar momentum and total radial momentum),  the properties of the  escaping particle, and the mass of the remaining particle. This makes the generalization to an arbitrary number of incident and ejected particles immediate.

\subsection{Initial conditions}

{While variation of $b_1$ (as in Eq. \ref{M2max1a}) may increase the CM energy, we find that, for the most efficient cases, this does not improve $E_3$ over $b_1=2$. For the sake of simplicity, we will therefore set $b_1=2$ from now on, unless we explicitly state otherwise.}
The other particle, denoted (2), possesses a subcritical angular momentum $b_{2}<2$. Under these conditions
the CM energy of such a collision diverges as the collision approaches the horizon of an extremal black hole  \cite{Banados+09}. Harada and Kimura   \cite{Harada2} have previously shown that the critical angular
momentum remains $L_{1}=2E_{1}$ for particles outside the equatorial
plane, as well. We therefore set $b_1 =2$ and $b_{2}<2$ in the following.

The direction of the radial momentum of each particle should be treated with special care, since not all scenarios allowed by the equations are physically reasonable. Particle (1) has a critical impact parameter, and can be either ingoing as it falls from infinity to the ergosphere, or outgoing if it has turned around the black hole before the collision. 
The sub-critical particle (2) cannot turn around, because there are no turning points within the ergosphere for $b<2$  -  hence it must be ingoing. Finally, 
particle (4) plunges into the black hole and as such is clearly ingoing  (though note that $\epsilon_{4}$ has no effect on Eq.  \ref{Eq:E3}). Considering these conditions, we have $\epsilon_2= \epsilon_4= - 1$ while $\epsilon_1$ can be $\pm 1$ . 

The case of an outgoing particle (1) was first considered by Schnittman   \cite{Schnittman14}, who examined a collision involving an initially infalling particle that has turned around the black hole. This makes the CM radial momentum less negative, and allows more energetic particles to escape to infinity. However, since this outgoing particle must necessarily be close to its turning point, its positive contribution to the radial momentum is limited, as by definition at the turning point $p^r = 0$.  Overall, this more elaborate configuration 
leads to an order-of-magnitude gain in the overall maximal energy when compared with the ingoing particle case.

The initial (right after collision) radial direction of the escaping particle is given by $\epsilon_3$.  Piran and Shaham   \cite{PiranShaham77} have shown that in the case of a collision with a total (center of mass) negative radial momentum, the efficiency is maximized for an initially ingoing particle. Indeed, examining Eq.  \ref{Eq:E3} we note that $\epsilon_3$ appears only in the denominator. 
As it has a positive prefactor (since $p_{tot} ^r<0$), the energy of the escaping particle is maximized for  $\epsilon_3= - 1$. This will set $\epsilon_2=\epsilon_3=\epsilon_4=-1$ and $\epsilon_1=\pm 1$.

 Since particle (3) is initially ingoing, it must have sufficient angular momentum to turn around the black hole, i.e.  $b_3\geq 2$.
 
\subsection{Maximizing the outgoing energy}
In the previous subsection we obtained a  general expression for the energy of a particle produced in a collision in a Kerr black hole's ergosphere. Here, we turn to our
main goal: the determination of the maximal possible energy of an escaping particle (3), under reasonable physical conditions, and the maximal possible efficiency of the process.  
The main challenge of this problem is in dealing with the large parameter phase space. We place these parameters in groups of diminishing importance:

\begin{itemize} 
\item{ (i) } The nature of the particles involved and in particular 
their masses $m_1,m_2,m_3, m_4$.  For our purposes, these define the physical process that takes place.

\item{ (ii)}  The orbital parameters of the incident particles  -  this group contains six parameters, $\{\sigma_I\}=\{E_{1},b_{1},\tilde p^{\theta}_{1},E_{2},b_{2},\tilde p^{\theta}_{2}\}$. 
\item{ (iii)} The coordinates of the collision point -  $r$ and $\theta$. We will sometimes be interested in finding the maximum efficiency for given  ($r$,$\theta$). 
\item{ (iv)} The outgoing particles  -  this group contains two parameters, $\{\sigma_E\}=\{b_{3},\tilde p_{3}^{\theta}\}$. This is the least interesting group  -   we maximize $E_3$ over this group under the condition that particle (3)  escapes to infinity.
\end{itemize}

Consider, for example, pair annihilation, defined by $E_{1}=m_{1}=E_{2}=m_{2}=1, m_{3}=m_{4}=0$. We begin with groups  (i)-(ii), by choosing certain values for the remaining, free $\{\sigma_I\}$  parameters. We then move to group  (iii), defining the collision point   $ (r,\theta) $ . We then find the set $\{\sigma_E\}$ that maximizes the efficiency at that particular point under the constraint that particle (3) escapes to infinity. To find the global maximum we  go over all possible values of $ (r,\theta) $, finding the appropriate values for $\{\sigma_E\}$ at every point, thus obtaining the maximum efficiency for those particular values of $\{\sigma_I\}$. We then move to group  (ii) and go over all values of the free $\{\sigma_I\}$ parameters, and repeat the previous steps for each value, thereby obtaining the \textit{global maximum efficiency} of the physical process.  

Once the parameters from groups  (ii) and  (iii) have been fixed, the maximal value of $E_{3}$ is obtained at   $b_3=\hat b$, 
$p_3 ^{\theta}=\hat p ^{\theta}$ for which   ${dE_{3}}/{db_{3}}=0$ , ${dE_{3}}/{dp_3 ^{\theta}}=0$ (these equations are typically solved numerically- though $\hat b$ can be calculated analytically in the case of a fully-equatorial collision, which will be defined shortly). However, $\hat b (r,\theta) $ defined this way may turn to be $< 2$, in which case particle (3) plunges into the black hole. This happens for   $r<r_{*}$ usually obeying $|r_{*} - 1| \ll1.$    
The derivative ${dE_{3}}/{db_{3}}$  is negative for $b_3=2$  -  hence, in this case,
the maximal energy for an escaping particle would be obtained for the critical value $b_3=2$. 
We define
\begin{equation} 
b_{*}\equiv \max (\hat b,2) ,
\label{Eq:b*}
\end{equation} 
and the maximal energy at a given point $ (r,\theta)$ is obtained by substitution of $b_3=b_*$
to Eq.  \ref{Eq:E3}.

For a given $\{\sigma_I\}$ and $\theta$, the maximum of the efficiency as a function of $r$ is always attained on the line $[1,r_{*}]$,  on which $b_{*}=2.$
Therefore substitution of $b_{3}=2$ maximizes the energy of the escaping particle (3) (see Fig. \ref{fig:Annihilation}). 
Note that this yields the \textit{global} maximum, after scanning over all values of  $r$. It does not give the maximal
energy for a general collision point,  $r$, since for $r> r_*$,  $b_{*} \ne 2$.

An important special case of the general solution is that of a fully-equatorial collision. We define an equatorial particle as a particle constrained to the equatorial plane($\theta=\frac{\pi}{2}$ with $p^\theta=0$, or equivalently $Q=0$), and a fully-equatorial collision as a collision wherein all particles are equatorial. Note that not every collision that takes place in the equatorial plane is necessarily fully-equatorial\footnote{Particles with a positive Carter constant will oscillate around the equatorial plane, and can therefore collide while crossing it.  While the collision takes place in the equatorial plane, neither the incident nor the ejected particles are confined to it, unless $Q_i=0$.}. Previous numerical calculations   \cite{Bejger+12,Schnittman14} considered only  fully-equatorial collisions, and we will compare our results to theirs where possible.

As we show later, in many cases a fully-equatorial collision maximizes the overall efficiency. The maximal energy for a fully-equatorial collision is
\begin{widetext}
\begin{eqnarray} 
&&E_{3,eq,max} (E_{tot}, L_{tot},m_4,r)= 
\left [{{E_{tot}^{2}}r^4 -  ({L_{tot}^{2} - E_{tot}^{2}}){r^{2}}+2{ (L_{tot} - E_{tot})^{2}}{r} - m_4 ^2 r^2 \Delta -  (\Sigma p_{tot}^{r})^2}  \right]
\\ \nonumber 
&& \bigg/ 2\left [{E_{tot} r^4+\Sigma^2 p_{tot}^{r} \nu (r,b_*) -  ({b_{*}L_{tot} - E_{tot}}){r^{2}}+2{ (b_{*} - 1) (L_{tot} - E_{tot})}{r}} \right]
  \ .
\label{Eq:E3maxeq}
\end{eqnarray} 
\end{widetext}
Bejger et al.     \cite{Bejger+12} suggested that due to symmetry considerations, the global upper limit of the efficiency will likely reside on the equatorial plane. While it's true that in all cases the global maximum resides on the equatorial plane, the most efficient collisions are not in all cases fully-equatorial. We will see an example of this in the case of pair annihilation.


\section{4. Specific Scenarios}
\label{sec:scenarios} 
In this section, we consider some specific physical scenarios. These include the pair annihilation of two massive particles, the scattering of a massless particle by a massive one  (e.g. Compton scattering)
and the  elastic scattering of two massive particles .

We find it useful to introduce a short-hand notation for the different physical scenarios considered. In this notation the first three characters represent the nature of particles (1), (2) and (3) respectively:
P for a photon, or M for a massive particle. The fourth character denotes the sign of the momentum of the critical particle  (1) :  ``+'' for outgoing, or  ``-''  for ingoing.  For example, $PMP-$  is the scattering of an ingoing critical photon by a subcritical massive particle, where the photon then escapes to infinity. Both particles are assumed to be initially infalling from infinity. Massive particles will be defined with $m=E=1$, and massless particles with $m=0$ and general $E$.

We will be interested in both the maximal energy and the maximal efficiency. We define the efficiency as the ratio of the energy of the escaping particle (3) to the total energy of the incident particles (1) and (2):
\begin{equation}
\eta \equiv \frac{E_{3}}{E_{tot}} \ .
\end{equation} 

After maximizing over all the orbital parameters of the colliding particles ($\{\sigma_I\}=\{E_{1},b_{1},\tilde p^{\theta}_{1},E_{2},b_{2},\tilde p^{\theta}_{2}\}$, denoted earlier as group (ii)) we find that for all processes considered here, the efficiency rises monotonically as $r$ approaches $1$. The maximal efficiency is attained in the $r\rightarrow 1$ limit, and is independent of particle (2)'s parameters. The efficiency may exhibit a discontinuity in $r=1$, as is the case in Compton scattering, which we discuss later.  

{Note that if we  limit the maximization to be  only over a sub-group of  these parameters (say, by setting $\tilde p_1 ^\theta=\tilde p_2 ^\theta=0$, as is the case in the fully-equatorial solutions), this behavior may change. In particular, the efficiency may attain a maximum at a finite distance away from the horizon. The location and value of the maximum may also depend on particle (2)'s parameters. These cases are more difficult to handle, and they always end up being less efficient than those where the maximum is attained at the $r\rightarrow 1$ limit. We will, however, discuss one such case, whose maximal efficiency was previously obtained numerically by \cite{Bejger+12}. }

\subsection{Fully-equatorial $MMP-$ collision}
We begin by considering a special case of the fully-equatorial annihilation of two massive ingoing particles 
that results in an escaping photon. Our goal in this subsection is to calculate the maximal efficiency of this special case.
By our short - hand notation, this is termed an $MMP-$ process. 

While, as we will see shortly,
the fully-equatorial case does not give rise to the global maximum of the $MMP-$ process,  it is important to derive this result here since it was previously calculated numerically   \cite{Bejger+12}. This will demonstrate the ability of our method to analytically derive this maximum in a very different way.

\begin{figure*}[t]
\includegraphics[width=85mm]{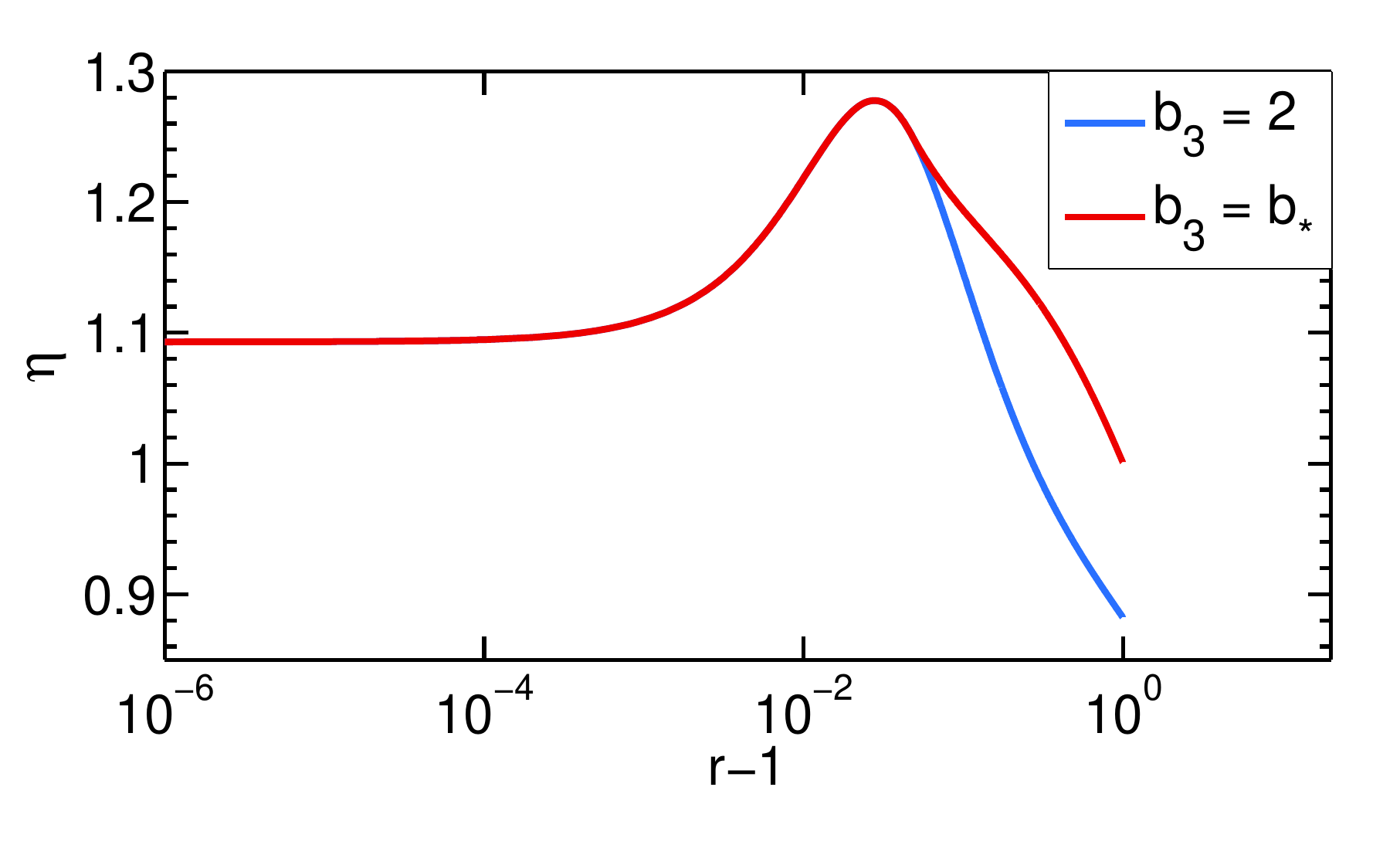}
\includegraphics[width=85mm]{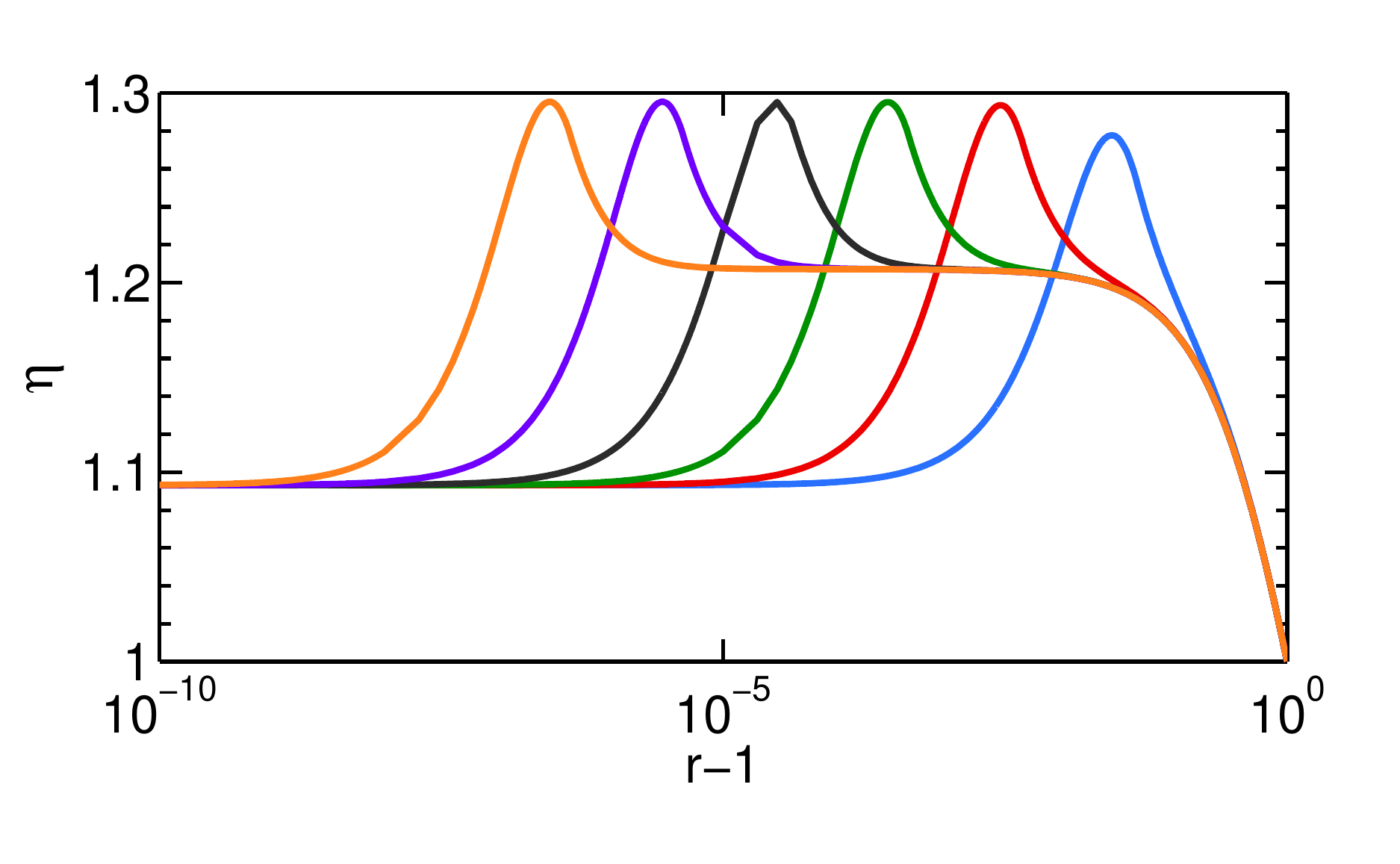}
\caption{Left: The efficiency of pair annihilation ($MMP-$) with $b_2=1.9$, obtained by using Eq.  \ref{Eq:b*}, as opposed to that obtained by setting $b_3 = 2$. The two functions  diverge only after attaining their maximum. For the purpose of finding that maximum, it is therefore sufficient to simply set $b_3 =2$. Right: The efficiencies of the fully-equatorial $MMP-$ process with $L_2$ logarithmically approaching 2 (after   \cite{Bejger+12}). Note how, while the maximum steadily approaches the horizon as $L_2$ approaches 2, the efficiency on the horizon itself is $\approx 1.09$ in all cases.}
\label{fig:Annihilation}
\end{figure*}

As always, particle (1) is critical, $L_{1}=b_{1}E_1 =2$, while $b_2 $ is sub-critical, satisfying $b_2 < 2$. 
Throughout this section, we will set $b_3=2$. While this is appropriate for calculating the \textit{global} maximal efficiency, the maximal efficiency at a given point is attained for the impact parameter given by  Eq.  \ref{Eq:b*}. This will only affect the tail of the function $\eta (r)$, toward $r=2$ (See Figure  \ref {fig:Annihilation}).

\begin{figure*}[t]
\includegraphics[width=85mm]{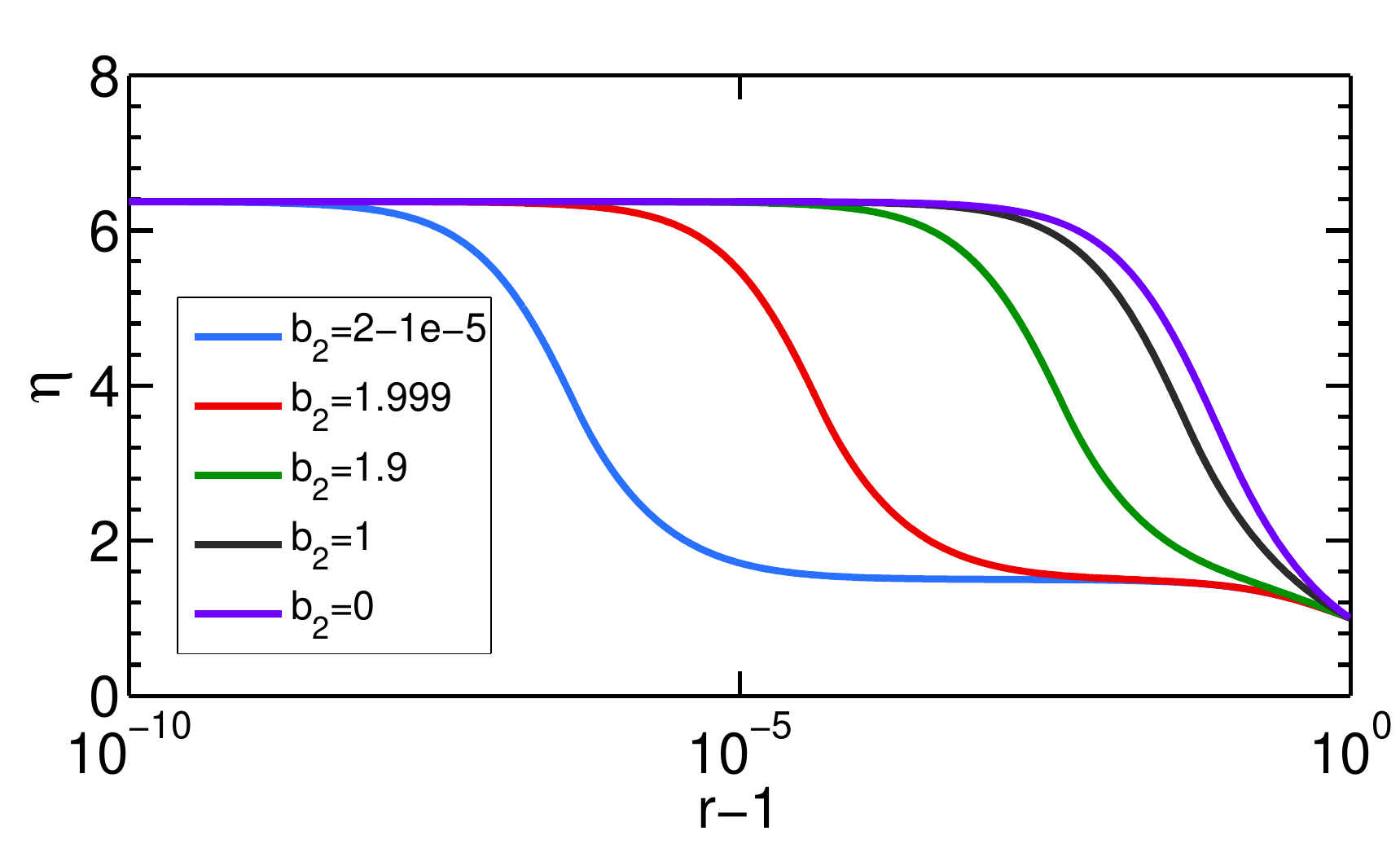}
\includegraphics[width=85mm]{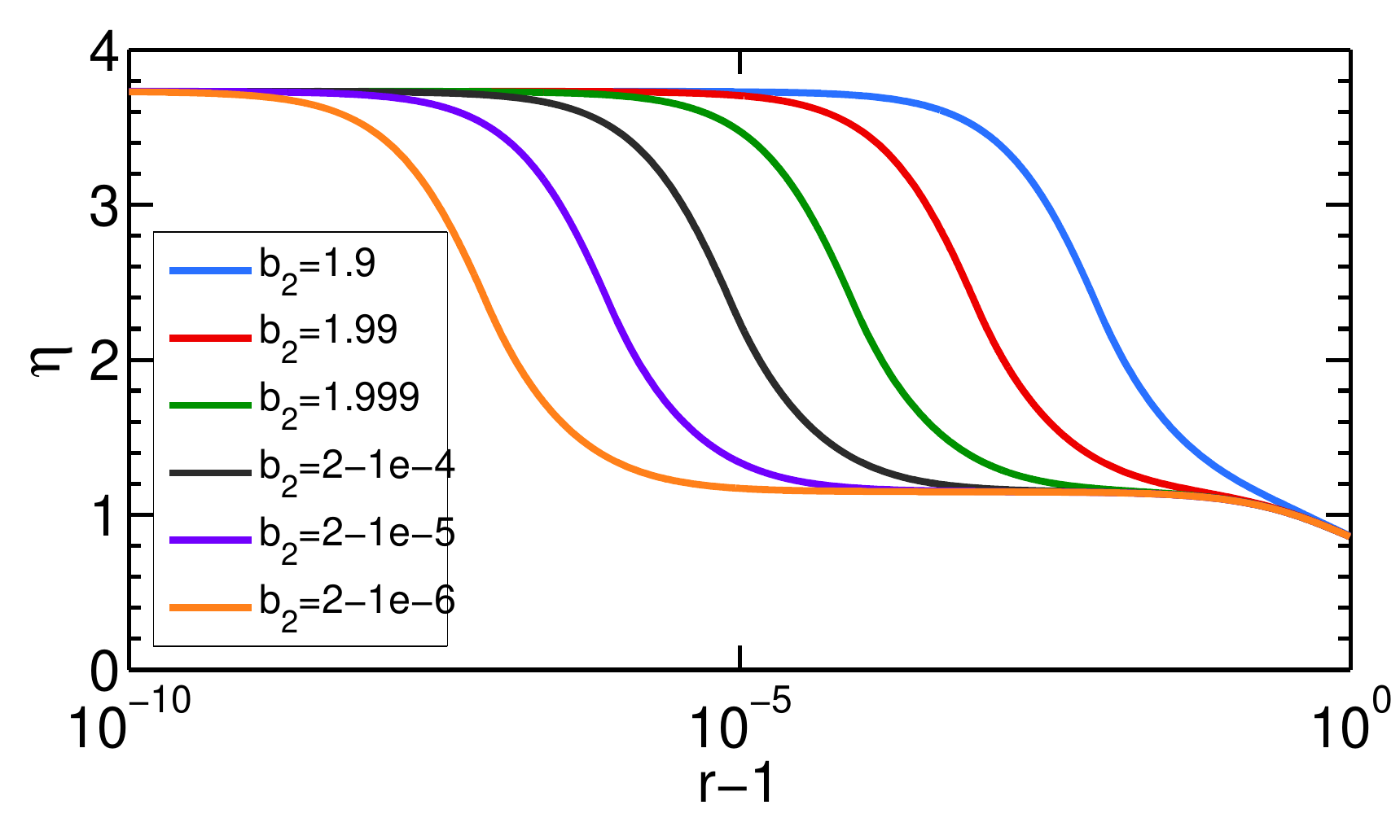}
\caption{ Left: The efficiency of the fully-equatorial $MMP+$ process. All efficiencies approach a maximum of $\approx 6.37$ in the $r\rightarrow 1$ limit, regardless of $b_2$. Right: The efficiency of the equatorial $MMP-$ process. We remove the restriction of a fully-equatorial collision, and particle (1) is given the polar momentum which maximizes the efficiency at every point $r$. Note how, once we're no longer limiting our parameters, the figure now resembles the $MMP+$ process, rather than the fully-equatorial $MMP-$ process.}
\label{fig:AnnihilationS}
\end{figure*}

Substituting these conditions into Eq.  \ref{Eq:E3maxeq}, we obtain
\begin{equation} 
\eta_{MMP-} (r,L_{2})=\frac{L_2 - r^2+A - 1}{2 (L_2 - r - r^2)+\sqrt{2r+4} (r+A - 1)} ,
\label{Eq:E3pair}\end{equation} 
with
\begin{equation}
A (r,L_2)=\sqrt{r^2 - \frac{L_2 ^2}{2} r+ (L_2 - 1)^2}.
\label{Eq:Adef}\end{equation}
 
Fig. \ref{fig:Annihilation} depicts $\eta (r)$
for different values of $L_{2}$ . Each curve attains a maximal
value at a different point $r=r_{max} (L_{2})$. As $L_{2}\rightarrow 2$, the maximal 
efficiency approaches a constant value  $\eta_{max}\approx1.295$ while 
$r_{max}\rightarrow 1$.

By naively taking the limit $r\rightarrow1$ for Eq.  \ref{Eq:E3pair}, we recover
Harada et al.'s   \cite{Harada} result of $\eta_{H}=[ (2+\sqrt{3}) (2 - \sqrt{2})]/{2}\approx1.093$,
which is independent of $L_{2}$. While $r_{max}\rightarrow 1 $  as $L_{2}\rightarrow 2$, 
one has to be careful in taking the limits. 
In order to obtain the real maximum they must
be taken concurrently.

This can be done by setting $L_{2}=2 - \xi,$ and $r_{max}=1+\delta\xi$, 
for some positive $\delta$, and taking the
limit $\xi\rightarrow$0 . This gives the constant term
\begin{equation} 
\eta_{max} (\delta)=\frac{4\delta -  (\delta+\sqrt{\delta^{2}+2\delta+\frac{1}{2}})^{2}+6\delta^{2}+\frac{1}{2}}{4\delta+12\delta^{2} - \sqrt{6}\delta (2\delta+2\sqrt{\delta^{2}+2\delta+\frac{1}{2}})}
\label{Eq:Emax???}\end{equation}

At $r_{max} (L_{2})$ the derivative
${\partial \eta}/{\partial r}=0$. Expanding 
${\partial \eta (L_{2}=2 - \xi,r)}/{\partial r}|_{r=1+\delta\xi}=0$,
we solve for the leading term in $\xi$ and obtain $\delta={1}/{\sqrt{12}}$, which yields
\begin{equation} 
\eta_{max}=\frac{1+\sqrt{3}+\sqrt{6}}{4}\approx1.295 \  .
\label{Eq:etamaxpair}\end{equation} 
This result was obtained numerically by \cite{Bejger+12}.

The fully-equatorial $MMP-$  collision is the only case we consider here where the maximum
efficiency $\eta_{max}$ is attained at a finite distance away from the horizon and is dependent on $L_2$.
This maximal efficiency is therefore the trickiest one to calculate. 

\begin{figure*}[t]
\includegraphics[width=85mm]{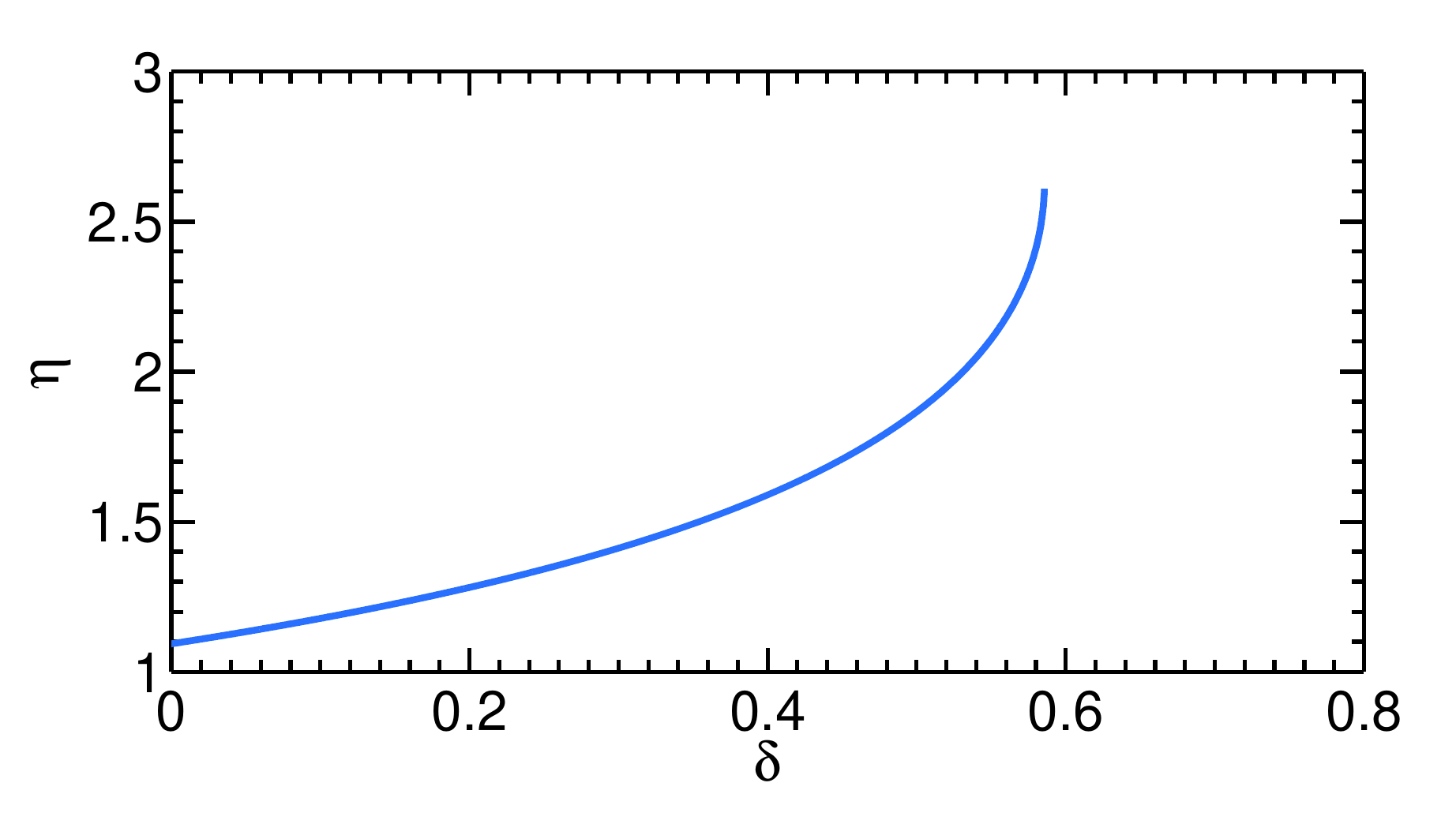}
\caption{ The efficiency of the fully-equatorial $MMP-$ process for $b_1=2+\delta (r-1)$ in the $r \rightarrow 1$ limit.  }
\label{fig:delta}
\end{figure*}

{We briefly mention that this maximum efficiency can be improved while still remaining fully-equatorial, by removing our restriction of $b_1=2$. Setting $b_1=2+\delta (r-1)$(see Eq. \ref{M2max1a}), we find that, for $r \rightarrow 1$ and $\delta=2-\sqrt{2}$, the efficiency approaches a maximum of $\eta_{max}=\frac{2+\sqrt{3}}{\sqrt{2}}\approx 2.63$. We plot $\eta_{max}$  as a function of $\delta$ in Fig. \ref{fig:delta}. As we will see, this efficiency is still smaller than that that can be attained by collisions that are not fully-equatorial.}

\subsection{Escaping massless particle}
For all following cases we find that the maximum is attained at the $r=1^+$ limit, and we can simply take the  $r\rightarrow 1$ limit of Eq. \ref{Eq:E3} to find the maximal efficiency. Note that there usually exists a discontinuity in the efficiency on the horizon itself. We present an example of this later in this section.

Here, we will immediately solve for the general non-equatorial collision. However, we find that the maximum is always attained on the equator(even though at times the collision which maximizes the efficiency in not fully-equatorial.)

Taking the limit $r\rightarrow 1$ (note that, when taking this limit, we must obviously set $b_3=2$) for  Eq.  \ref{Eq:E3},  we obtain
\begin{widetext}
\begin{equation}
E_{3, r\rightarrow 1} (E_1 , m_1 , \tilde p^\theta _1 , \tilde p_3 ^\theta , \sin\theta)=\frac{2 E_1 \sin\theta +\epsilon_1 \sqrt{ (m_1 ^2 - E_1 ^2) \sin^4\theta  + [8 E_1 ^2  - 2 m_1 ^2-E_1 ^2 (\tilde p_1^\theta)^2] \sin^2\theta - 4 E_1 ^2}}{2 \sin\theta - \sqrt{ - \sin\theta^4 + [8-(\tilde p_3 ^\theta)^2]\sin^2\theta  - 4}} \ .
\label{Eq:Efor1}\end{equation} 
\end{widetext}

Remarkably, this result is completely independent of particle (2)'s parameters.

Eq. \ref{Eq:Efor1} will clearly be optimized by choosing $\tilde p_3 ^\theta=0$, which we will always impose when discussing the $r\rightarrow 1$ limit. For a collision taking place in the equatorial plane, the maximal energy will always be obtained for an equatorial particle (3). Setting the proper parameters for the $MMP-$ case, we find
\begin{eqnarray}
&&\eta_{MMP-, r\rightarrow 1} (\sin\theta,\tilde p_1 ^\theta)=\nonumber \\
&&\frac{2\sin\theta - \sqrt{ - \sin^2\theta  (\tilde p_1 ^\theta)^2+6 \sin^2\theta - 4}}{2 (2\sin\theta - \sqrt{ - \sin^4\theta+8\sin^2\theta - 4})}
\end{eqnarray}
This means that the result is optimized not for $\tilde p_1 ^\theta=0$, but for $ (\tilde p_1 ^\theta)^2=({6\sin^2\theta - 4})/({\sin^2\theta})\equiv  (\tilde p_{1, max} ^\theta)^2   (\sin\theta)$. 

Generally, after taking the limit $r\rightarrow 1$ we find that collisions with an ingoing particle (1) are optimized by the maximal possible value of $\tilde p_1 ^\theta$ (which will eliminate the square root in Eq.  \ref{Eq:Efor1}'s numerator), while those with an outgoing particle (1) are optimized by $\tilde p_1 ^\theta=0$. This is in line with Eq. \ref{Eq:Mcmlimit}-\ref{M2max2a}, where a positive $Q_1$ raises the CM energy of a collision involving two ingoing particles, and lowers the CM energy of a collision involving one outgoing particle.

The resulting efficiency for the $MMP-$ case is a monotonically increasing function of $\sin\theta$, and is defined for $\sqrt{\frac{2}{3}}\leq \sin\theta\leq 1$. This function attains a maximal value of  $\eta_{max}=2+\sqrt{3}\approx 3.73$ at $\sin\theta=1$, which is the global maximum for the $MMP-$  case.

The maximal energies and efficiencies for all other cases can also be derived from Eq. \ref{Eq:Efor1}. The results are summarized in table $1$. In all cases the maximal efficiency is attained on the equatorial plane, and Eq. \ref{Eq:Efor1} necessitates us to set $b_1=b_3=2$.

This  last result, $b_1=b_3=2$,   may at first seem paradoxical for the $PMP+$ case,
where the incident and ejected particles are both massless. 
Both are equatorial with $b=2$. For an equatorial massless particle, the impact parameter $b$ completely specifies the geodesic: $b_1=b_3=2$ therefore  seemingly implies both move on the same trajectory. Namely, no collision took place in the CM frame, and particle (3) should emerge from the collision with energy $E_1$. However, there is an additional degree of freedom: the direction of motion of the massless particle in the radial direction. 

\begin{figure*}[t]
\includegraphics[width=110mm]{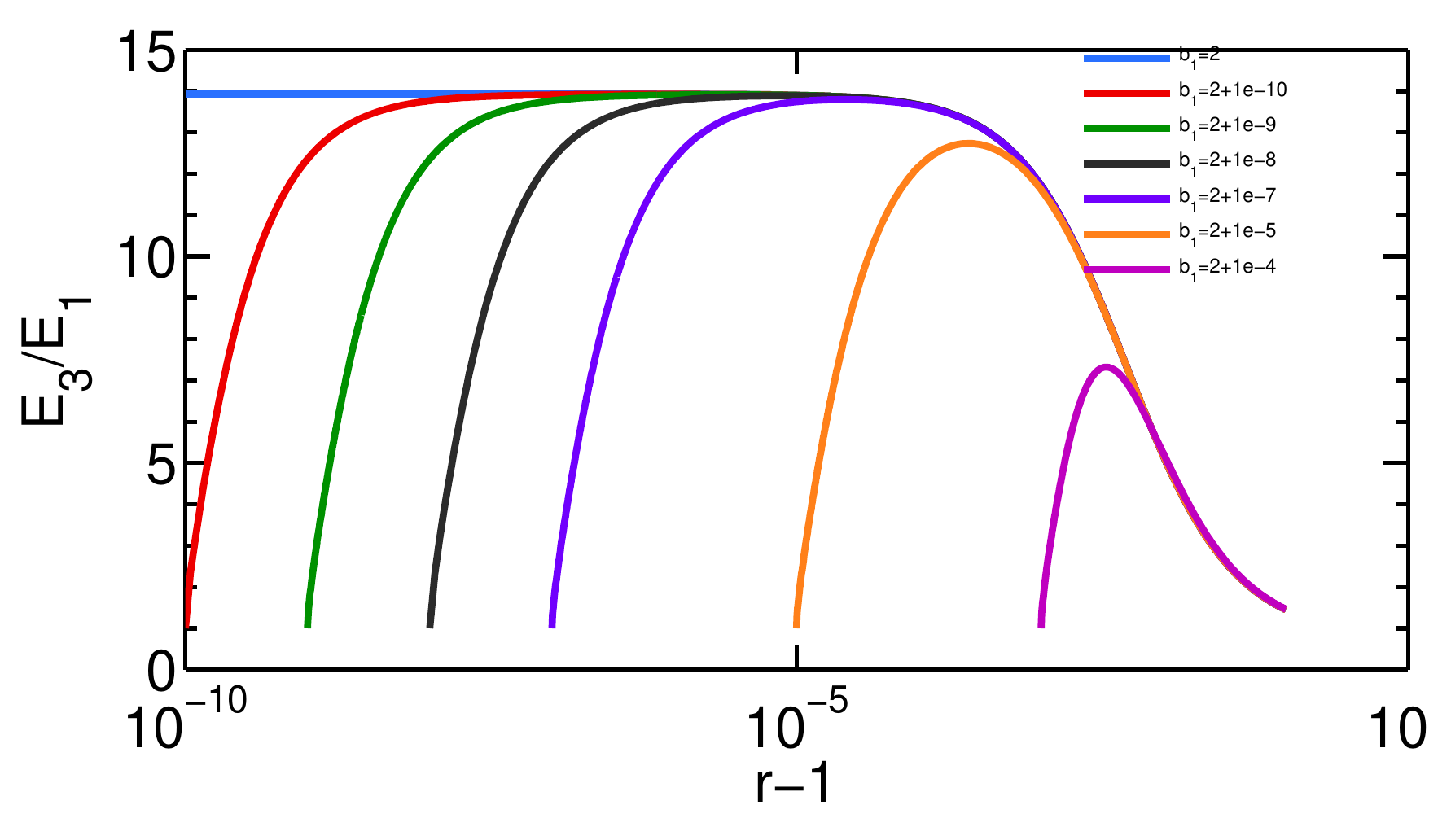}
\caption{The energy $E_3$ obtained by the fully-equatorial $PMP+$ process is plotted with $b_1=b_3=2+\epsilon$ and $E_1=1 , b_2=1.9$. }
\label{fig:Asymp}
\end{figure*}

To gain some insight into the situation, we plot the $PMP+$ process with $b_1=b_3=2+\epsilon$, for $\epsilon<1$(Fig. \ref{fig:Asymp}). The energy, $E_3$,  increases as $r$ decreases. It reaches a plateau at some  maximal value, and then decreases monotonically, arriving at the value $E_1$  on the radial turning point, defined as $r_1=1+\epsilon$. The relative width of this descent from the maximal value down to $E_1$ decreases,  and the width of the plateau increases, as $\epsilon\rightarrow 0$ . For $b_1=b_3=2$, we therefore expect that $E_3$ will sustain its plateau for all $r>1$, and drop discontinuously to $E_3=E_1$ at $r=1$. 

Even though both particles have the same impact parameter and therefore move on the same geodesic, the incident particle is outgoing, while the ejected particle is initially ingoing. The particle's energy  increases while its radial velocity is reversed. At the radial turning point $p^r=0$ by definition, and so no collision actually takes place and $E_3=E_1$.

\hfill \break

\begin{widetext}
\begin{center}
    \begin{tabular}{|c| l | l | l | l |  p{4cm}| }
    \hline
   & Maximal energy & Maximal efficiency  \\ \hline
    MMP -&  $2(2+\sqrt{3})$ & $2+\sqrt{3}\approx 3.73$  \\ \hline
    MMP+& $(2+\sqrt{3})(2+\sqrt{2})$& ${ (2+\sqrt{3}) (2+\sqrt{2})}/{2} \approx 6.37^\dagger $\\ \hline
    PMP - & $2(2+\sqrt{3}) E_1 $ & $2 (2+\sqrt{3}) \approx 7.46 ^* $ \\\hline
    PMP+& $(2+\sqrt{3})^2 E_1$& $ (2+\sqrt{3})^2 \approx 13.92^{*\dagger} $ \\\hline
    MPP - & $2(2+\sqrt{3})$&  $2 (2+\sqrt{3}) \approx 7.46^{**} $ \\\hline
    MPP+&  $(2+\sqrt{3})(2+\sqrt{2})$&  $ (2+\sqrt{2}) (2+\sqrt{3}) \approx 12.74^{**} $ \\\hline
    \end{tabular}
    \break
 \end{center}
    {Table 1: Maximal energies and efficiencies for the different cases considered for an escaping photon. Fully-equatorial collisions maximize the ``$+$'' cases, while the ``$-$'' cases are maximized by a particle (1) with the maximal allowed polar momentum.
\hfill\break
$^\dagger$- Obtained numerically by \cite{Schnittman14}
\hfill\break
$^*$- The maximal efficiency is attained for $E_1\rightarrow \infty$
\hfill\break
$^{**}$- The maximal efficiency is attained for $E_2\rightarrow 0$}
\end{widetext}

\begin{figure*}[t]
\includegraphics[width=110mm]{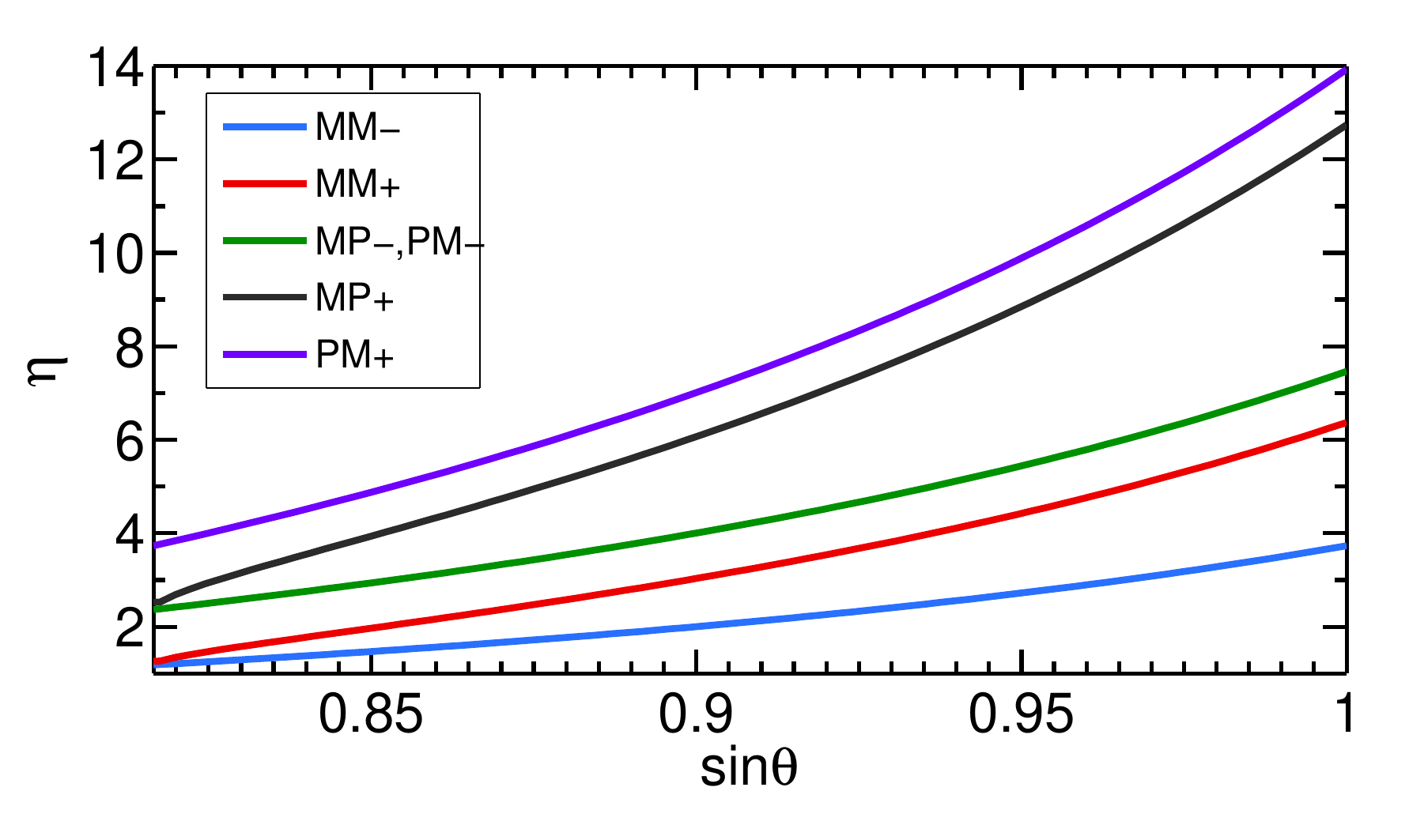}
\caption{The maximal efficiencies for all six cases for a massless escaping particle.}
\label{fig:massless}
\end{figure*}

Figure  \ref{fig:massless} shows the maximum efficiencies of all six cases as a function of $\sin\theta$. The efficiency is in all cases a monotonically increasing function of $\sin\theta$, and the maximum in all cases therefore resides on the equatorial plane.

 Note that only the ``$+$'' collisions are fully-equatorial, while the ``$-$'' collisions are maximized by a particle (1) with the maximal allowed polar momentum. In the $PMP-$ case, in particular, the incident photon (1) possesses the maximal allowed polar momentum, while the ejected photon (3) is equatorial with $\tilde p_3 ^\theta =0$- thus, they don't even move on the same geodesic.

The $PMP$ cases, where the critical particle is massless, can occur on a wider range of angles, $\sqrt{3} - 1 \leq \sin\theta \leq 1$. Since the efficiency decreases with the angular distance from the equator, collisions in this additional area will not be very efficient.

It should be mentioned that collisional Penrose processes can still be possible for smaller values of $\sin\theta$, provided we remove our restriction of $b_1 = 2$. This will necessarily shift the innermost turning point away from the horizon, and the efficiency is expected to drop rapidly. These cases are therefore irrelevant for estimates of maximal efficiencies.

\subsection{Escaping massive particle}

In previous sections, we examined the case where the escaping particle, (3), is massless. Here we will obtain the energy of an escaping particle with a general mass, parametrized by $\alpha_3 \equiv {m_3}/{E_3}$. We will use this more generalized formula to obtain the maximum efficiency for the case of an elastic scattering (defined by $m_1=E_1=m_2=E_2=m_3=m_4=1$), as well as for all other cases where the escaping particle is the massive one.

The energy $E_3$ in Eq.  \ref{Eq:E3} is obtained by solving Eq.  \ref{Eq:prC} -  which, for a massless particle (3), is simply a linear equation in $E_3$:
\begin{equation}
B E_3 - C=0
\end{equation}
with $B$ and $C$ being, respectively, the denominator and numerator of Eq.  \ref{Eq:E3}. When particle (3) is massive, the energy $E_3$ becomes the solution to a quadratic equation:
\begin{equation}
A E_3 ^2 +B E_3  - C =0
\end{equation}
with $A =  - \Delta \Sigma\alpha_3 ^2$. Note that the parametrized mass $\alpha_3$ also appears in the factor $\nu_3$, as defined in Eq. \ref{Eq:pr3q}.

One can then verify, by taking the derivative ${d E_3}/{d \alpha_3}$ and directly plugging in $\alpha_3=0$, that a massless particle will always be an extremum of the efficiency. Surprisingly, $\alpha_3=0$ will sometimes minimize the efficiency (particularly for $r\approx 2$, i.e. far from the horizon), but it is always a maximum for $r\rightarrow 1$, which is the region we're interested in. Taking the limit $r \rightarrow 1$, we find the generalized version of Eq.  \ref{Eq:Efor1}

\begin{widetext}
\begin{equation}
E_{3, r\rightarrow 1} (E_1 , m_1 , \tilde p^\theta _1 , sin\theta,\alpha_3)=\frac{2 E_1 \sin\theta +\epsilon_1 \sqrt{ (m_1 ^2 - E_1 ^2) \sin^4 \theta  + [ 8 E_1 ^2 - 2 m_1 ^2- E_1 ^2   (\tilde p_1 ^\theta) ^2]\sin^2 \theta - 4 E_1 ^2}}{2 \sin\theta - \sqrt{8 \sin^2\theta - \sin^4 \theta  - 4 - \alpha_3 ^2  (2 \sin^2 \theta - \sin^4\theta)}}
\label{Eq:Emfor1}\end{equation} 
\end{widetext}
Plugging in the numbers for an elastic collision, we find

\begin{equation}
E_{3,r\rightarrow 1}  (\tilde p^\theta _1 , \alpha_3)=\frac{2+\epsilon_1 \sqrt{2 -  (\tilde p^\theta _1) ^2}}{2 - \sqrt{3 - \alpha_3 ^2}}
\end{equation}
We now enforce the elastic collision by demanding that 
\begin{equation}
m_3=\alpha_3 E_3=1 .
\label{Eq:mass}
\end{equation}
 Solving Eq. \ref{Eq:mass} for $\alpha_3$, we can then find the corresponding maximal efficiencies for both the ingoing ($\eta_{max}=({4+\sqrt{11}})/{2}\approx 3.66$) and the outgoing ($\eta_{max}=({7+4\sqrt{2}})/{2}\approx 6.32$) cases.

We can similarly formulate the results of the previous section for Compton scattering, for the case where the escaping particle is the massive one. These are summarized in table $2$. 

More interestingly, we can calculate for each case the maximal mass of an escaping particle, by maximizing $m_3$ as given by Eq. \ref{Eq:mass}. It is clearly bounded from above by $E_{max}$, but will not necessarily reach this limit. Furthermore, this maximal mass is not always attained for the same conditions under which the maximal energy is attained. 
\begin{figure*}[t]
\includegraphics[width=110mm]{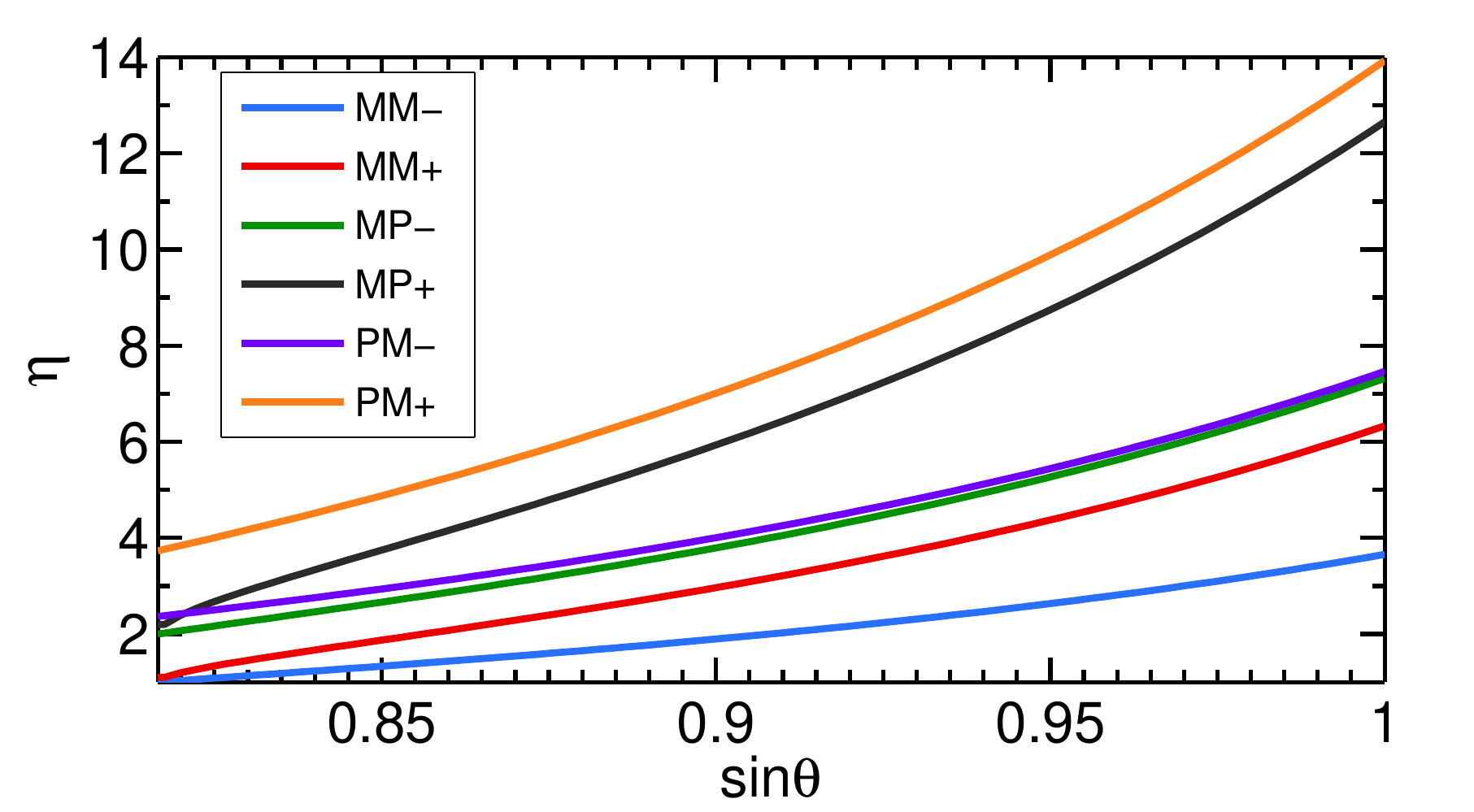}
\caption{The maximum efficiencies for all six cases for a massive escaping particle (3).}
\label{fig:massive}
\end{figure*}

{Note how both $PMM$ cases define a minimal $E_1$. This is the minimal energy necessary for a critical particle to emerge with $m_3=1$.}

The upper bounds for the MPM and MMM cases are, as expected, lower than their massless counterparts. While the $E_1\rightarrow \infty$ upper bounds for the PMM cases are equal to their massless counterparts , they are always lower -  this difference simply goes to zero when taking the limit.

Finally, using Eq. \ref{Eq:Emfor1} we can explore the effect of the mass of particle $(3)$ on the maximal energy. 
This mass, $m_3$, appears (via $\alpha_3$) only in the denominator. It   has to be minimal for the denominator to be smallest. All other factors in the denominator are of order unity and hence we expect that $E_3$ will decreases significantly only once $\alpha_3$ is of order unity, i.e. when $m_3 \approx E_3$. For example for an equatorial collision ($\theta = \pi/2$), when $\alpha_3 =1$, its maximal value for an escaping particle, $E_3$ decreases to about 40\% of its maximal value attained for $\alpha_3 = 0$. 

\begin{widetext}
\begin{center}
    \begin{tabular}{ | l | l | l | l |p{4cm} |}
    \hline
                 & Maximal energy & Maximal efficiency   & Maximal mass \\\hline
    MMM -  &$4+\sqrt{11}$ & $({4+\sqrt{11}})/{2}\approx 3.66$ & $2\sqrt{3}\approx 3.46$\\\hline
    MMM+ & $7+4\sqrt{2}$& $({7+4\sqrt{2}})/{2}\approx  6.32$  & $\sqrt{3}(2+\sqrt{2})\approx 5.91$ \\\hline
    PMM -  &$4 E_1+\sqrt{(12 E_1 ^2-1)}$ & $2 (2+\sqrt{3})\approx  7.46^{*} $& $2\sqrt{3} E1\approx 3.46 E_1$\\\hline
    PMM+ &$2(2+\sqrt{3}) E_1+\sqrt{[3(2+\sqrt{3})^2 E_1 ^2-1]} $ & $ (2+\sqrt{3})^2 \approx  13.92^{*}$  & $\sqrt{3}(2+\sqrt{3}) E_1\approx 6.46E_1$ \\\hline
    MPM -  & $4+\sqrt{11}$ & $4+\sqrt{11}\approx  7.32^{**}$& $2\sqrt{3} \approx 3.46$\\\hline
    MPM+ &  $7+4\sqrt{2}$& $7+4\sqrt{2}\approx  12.66 ^{**}$& $\sqrt{3}(2+\sqrt{2})\approx 5.91$ \\\hline
    \end{tabular}
    \\
\end{center}    Table 2: Maximal efficiencies for the different cases for a massive escaping particle. Note that the maximal mass is not attained at the same collision as the maximal energy. 
\hfill\break
$^*$- The maximal efficiency is attained for $E_1\rightarrow \infty$
\hfill\break
$^{**}$- The maximal efficiency is attained for $E_2\rightarrow 0$

\end{widetext}

\section{5. Discussion}
\label{sec:conclusions}

We have obtained a general formula for  the maximal energy  of a particle that escapes to infinity following a collisional Penrose process \cite{Piran+75}. We have considered a maximal ($a=1$) Kerr black hole, for  which the highest energies are attainable. We have  applied this formula to several cases, and in particular to the annihilation of two particles falling from rest at infinity and to the  scattering of a massless particle by a massive one falling from rest at infinity. 
In all cases, the maximal energy is obtained when one of the particles is ``critical" , namely it has a critical ratio of angular momentum to energy $b\equiv L/E = 2$, and hence a turning point on $r=1$.  The energy of the escaping particle, (3), in the $r \rightarrow 1$ limit, depends only on the properties of the critical particle (1). 

Even though the CM energy of these near-horizon collisions diverges, the maximal energy  of the escaping particles is always finite and rather modest. 
For the annihilation of ingoing particles, the maximal energy is $\sim 7.46 m_1$ for the case of an escaping photon and slightly lower ($\sim 7.32 m_1$) for an escaping massive particle.  
The corresponding energy gains, $E_3-2 m_1$, are larger than those obtained by \cite{Bejger+12} for fully- equatorial collisions by a factor of 9, and are almost comparable to those obtained by \cite{Schnittman14} for the annihilation of an ingoing particle with a particle that has turned around the black hole. 

Interestingly, even though the maximal energy is obtained for collisions that take place in the equatorial plane and the escaping particle is always constrained to the equatorial plane, these values are larger than those obtained for a fully-equatorial collisions, in which the colliding particles also move in the equatorial plane. This happens because, in this case, when the colliding particles are both ingoing and are not constrained to the equatorial plane, the maximal CM energy is larger(see Eq. \ref{M2max1a} ) than that obtained if both particles move in the equatorial plane.  
  
If the critical particle has turned around the black hole and is outgoing, the maximal energy is larger: $\sim 12.74 m_1$ if the escaping particle is a photon, and slightly lower $\sim 12.64 m_1$, if it is a massive particle.  In this case, the CM energy is maximal when the collision is fully-equatorial(Eq. \ref{M2max2a}), and hence our result agrees with the fully-equatorial numerical result obtained  earlier by  \cite{Schnittman14}. 

The closest analytical results are due to Harada et al. \cite{Harada}, who used a power-series expansion of the radial momentum equation (Eq. \ref{Eq:prC}) to obtain the efficiency of a fully-equatorial collision taking place directly on the horizon, where particle (1) is ingoing. Under these assumptions, our Eq. \ref{Eq:Efor1} reduces to their 4.12. Since the fully-equatorial case maximizes only the ``$+$'' cases, which were not considered by \cite{Harada}, our upper bounds do not coincide with theirs.

In the case where the escaping particle is massive, an additional kinematic condition exists: the mass of the particle must be smaller than the CM energy of the system. However, a more important bound is that to reach infinity $m_3 \le E_3$. This poses a strong limit on the maximal masses of the escaping particles: $\sim 3.46 m_1$ and $\sim 5.91 m_1$ for an ingoing and outgoing critical incident particle, respectively. This implies that, while  the CM energy diverges when the colliding particles have suitable momenta and the collision takes place near the horizon, even if an exotic massive particle will be produced in such a collision, it won't be able to escape and reach an observer at infinity.

The highest efficiency of the collisional Penrose process, among the processes that we considered,  is obtained for the scattering of an outgoing critical massless particle by a massive particle. The escaping particle can be either massless or massive. In the limit where the  energy of the incident massless particle is very large, the efficiency of the process approaches
\begin{equation}
\eta_{PMP+}=\eta_{PMM+}=(2+\sqrt{3})^2\approx 13.92 \ .
\end{equation}
This is an upper bound on any collisional Penrose process involving particles that have originated at infinity\footnote{To be precise, the limit applies to the case where the massive particle falls from rest at infinity}.

Our method can be used to solve more general problems than those considered
here. Specifically, we focused on the global maximal energy of the escaping particle, and the special case of two particles falling from infinity.
We stress that there is nothing special about the case of two - particle collisions. Eq.
 \ref{Eq:E3} depends only on the total energy and momenta of the incident particles. As such, it can be used for any combination of incident particles, and even
for the original Penrose process in which a single particle disintegrates within the ergosphere.

More generally, we can consider different initial momenta than those we considered here and, while our discussion focused on the case that the incident particles fall from rest from infinity, one can use our formalism for more general cases. 
We simply follow the same procedure, maximizing $E_3$ over different choices of $b_3$ and $\tilde p^\theta_3$ for a given set of parameters. This would allow us to obtain the local maxima of the energy for specific collisions.  For example, we have used this method in   \cite{LP15} to estimate the maximal efficiency in a specific collision suggested by Berti et al.   \cite{Berti+14}.

When considering more general initial conditions for the incident particle one should, however, proceed with caution. Not all parameters that are mathematically allowed at a given point ($r,\theta$)  are physically reasonable. Specifically, there is a region in the phase space of allowed momenta in which the particles must have emerged from the black hole (or generated somehow very close to the horizon on an outgoing orbit). 

This was first considered by Berti et al.  \cite{Berti+14}, who treated the case where both incident particles are sub-critical, and yet one of them is outgoing. Since there are no turning points within the ergosphere for such particles, their radial momentum doesn't have to be small and, unsurprisingly, such collisions can result in diverging energies. However, the outgoing sub-critical particle cannot have originated at infinity(see Fig. \ref{fig:b_a1}). As it cannot have emerged from the black hole, it must therefore have been produced within the ergosphere, very near to the black hole's horizon.  

Berti et al. suggested that this particle is produced in a prior collision of two particles infalling from infinity. While kinematically allowed, this now changes the overall energy budget. 
When the energy of these infalling particles is taken into account in the estimates of the overall efficiency of the process,  the resulting efficiency reduces back to Schnittman-like levels. Thus, even though this is an unlikely configuration, involving two fine-tuned collisions that take place infinitely close to the black hole's horizon, this does not result in any extra gain in energy or efficiency when compared with simpler collisions.

To conclude, we discuss briefly the astrophysical and ``technological" implications of our findings. We have shown that under extreme conditions ($a\rightarrow 1\; , \  r\rightarrow 1^+\; ,\  b_1\rightarrow  2^+$) a  collision can result
in a  particle that escapes to infinity with energy of a factor of a few (up to about an order of magnitude) larger than the energies of the incident particles. This is clearly a Penrose process, as the energy must come from the rotational energy of the black hole. While this is a significant energy gain, this result rules out two ideas that gained (unjustified) popularity recently. 

The first is that, with the possibility of diverging CM energies in such collisions, Supermassive black holes can accelerate, in this way, the observed Ultra High Energy Cosmic Rays (UHECRs) that are detected at up to $10^{20}$eV. There is no way that a proton or a nuclei will be accelerated to such energies and escape from the vicinity of the black hole.

 A second idea that was put forward was that, with  diverging CM energies, such collisions could produce exotic massive particles that cannot be produced otherwise, and as such could be used to explore physics that cannot otherwise be 
explored. However, as one can see from the maximal rest masses derived (see Table 2), such particles, if produced, won't be able to escape to an observer at infinity. 

In this work we focused on the maximal possible energies in different collisions. These values are attained for an idealized situation of a maximally rotating black hole and for fine-tuned collisions taking place infinitely close to the horizon. Any deviation from these idealized condition will drastically reduce the resulting energies. 

We didn't address the interesting astrophysical question of the likelihood of such collisions to take place. Still, examination of 
Figs. \ref{fig:Annihilation},  \ref{fig:AnnihilationS} and \ref{fig:Asymp} reveals that to be effective, such collisions must be extremely close to the event horizon.
This implies that the effective geometrical cross section of the regions in which such collisions would take place is extremely small. Furthermore, 
the angular momentum of one of the particles must be extremely close to the critical value. Thus, the chances of having relevant collisions are extremely small. 

A final issue that we don't address at all is 
the differential cross section - that is, the fraction of the angular phase space of the scattered particle in which energy gain is significant and the particle escapes. We suspect that the combination of all these constraints makes it unlikely that these collisions play an important role in either particle acceleration, particle production or energy extraction in  either astrophysical or ``technological"   systems. 

This research was supported by a grant from ISA the Israel Space Agency and by the I-CORE Program of the Planning and Budgeting Committee
and The Israel Science Foundation


\def\apj{Astrophys.\ J.}
\def\nat{Nature}
\def\apjl{Astrophys.\ J. Lett.}
\def\apj{Astrophys.\ J.}
\def\aap{Astron.\ Astrophys.}
\def\prd{Phys. Rev. D}
\def\physrep{Phys.\ Rep.}
\def\mnras{Month. Not. RAS }
\def\araa{Annual Rev. Astron. \& Astrophys.}
\def\aapr{Astron. \& Astrophys. Rev.}
\def\aj{Astronom. J.}
\def\jcap{JCAP}

%



\end{document}